\documentclass[prx,aps, unsortedaddress, superscriptaddress]{revtex4-1}
\pdfoutput=1
\usepackage[english]{babel}
\usepackage{amsmath,amssymb,graphicx,enumerate,hyperref}

\catcode`\>=\active \def>{
\fontencoding{T1}\selectfont\symbol{62}\fontencoding{\encodingdefault}}
\newcommand{\tmem}[1]{{\em #1\/}}
\newcommand{\tmnote}[1]{\thanks{\textit{Note:} #1}}
\newcommand{\tmop}[1]{\ensuremath{\operatorname{#1}}}
\newcommand{\tmtextit}[1]{{\itshape{#1}}}
\newcommand{\tmverbatim}[1]{{\ttfamily{#1}}}
\newenvironment{enumerateroman}{\begin{enumerate}[i.] }{\end{enumerate}}
\newenvironment{quoteenv}{\begin{quote} }{\end{quote}}

\begin{document}

\title{Absence of a Dissipative Quantum Phase Transition in Josephson
Junctions}

\author{A. Murani}
\affiliation{Universit{\'e} \ Paris-Saclay, \ CEA, \ CNRS, \ SPEC\\
91191 \ Gif-sur-Yvette \ Cedex, France}

\author{ N. Bourlet}
\affiliation{Universit{\'e} \ Paris-Saclay, \ CEA, \ CNRS, \ SPEC\\
91191 \ Gif-sur-Yvette \ Cedex, France}

\author{H. le Sueur}
\affiliation{Universit{\'e} \ Paris-Saclay, \ CEA, \ CNRS, \ SPEC\\
91191 \ Gif-sur-Yvette \ Cedex, France}

\author{F. Portier}
\affiliation{Universit{\'e} \ Paris-Saclay, \ CEA, \ CNRS, \ SPEC\\
91191 \ Gif-sur-Yvette \ Cedex, France}

\author{C. Altimiras}
\affiliation{Universit{\'e} \ Paris-Saclay, \ CEA, \ CNRS, \ SPEC\\
91191 \ Gif-sur-Yvette \ Cedex, France}

\author{D. Esteve}
\affiliation{Universit{\'e} \ Paris-Saclay, \ CEA, \ CNRS, \ SPEC\\
91191 \ Gif-sur-Yvette \ Cedex, France}

\author{H. Grabert}
\affiliation{Physikalisches Institut, Universit{\"a}t Freiburg,\\
Hermann-Herder-Stra{\ss}e 3, 79104 Freiburg, Germany}

\author{J. Stockburger}
\affiliation{Institute for Complex Quantum Systems and IQST,\\
University of Ulm, 89069 Ulm, Germany}

\author{J. Ankerhold}
\affiliation{Institute for Complex Quantum Systems and IQST,\\
University of Ulm, 89069 Ulm, Germany}

\author{P. Joyez}
\tmnote{ email: philippe.joyez@cea.fr}
\affiliation{Universit{\'e} \ Paris-Saclay, \ CEA, \ CNRS, \ SPEC\\
91191 \ Gif-sur-Yvette \ Cedex, France}

\date{March 13, 2020}

\begin{abstract}
  Half a century after its discovery, the Josephson junction has become the
  most important nonlinear quantum electronic component at our disposal. It
  has helped reshape the SI system around quantum effects and is used in
  scores of quantum devices. By itself, the use of Josephson junctions in the
  volt metrology seems to imply an exquisite understanding of the component in
  every aspect. Yet, surprisingly, there have been long-standing subtle issues
  regarding the modeling of the interaction of a junction with its
  electromagnetic environment. Here, we find that a Josephson junction
  connected to a resistor does not become insulating beyond a given value of
  the resistance due to a dissipative quantum phase transition, as is commonly
  believed. Our work clarifies how this key quantum component behaves in the
  presence of a dissipative environment and provides a comprehensive and
  consistent picture, notably regarding the treatment of its phase.
\end{abstract}

{\maketitle}

\section{Introduction}

\begin{figure}[hbt]
  \resizebox{8.5cm}{!}{\includegraphics{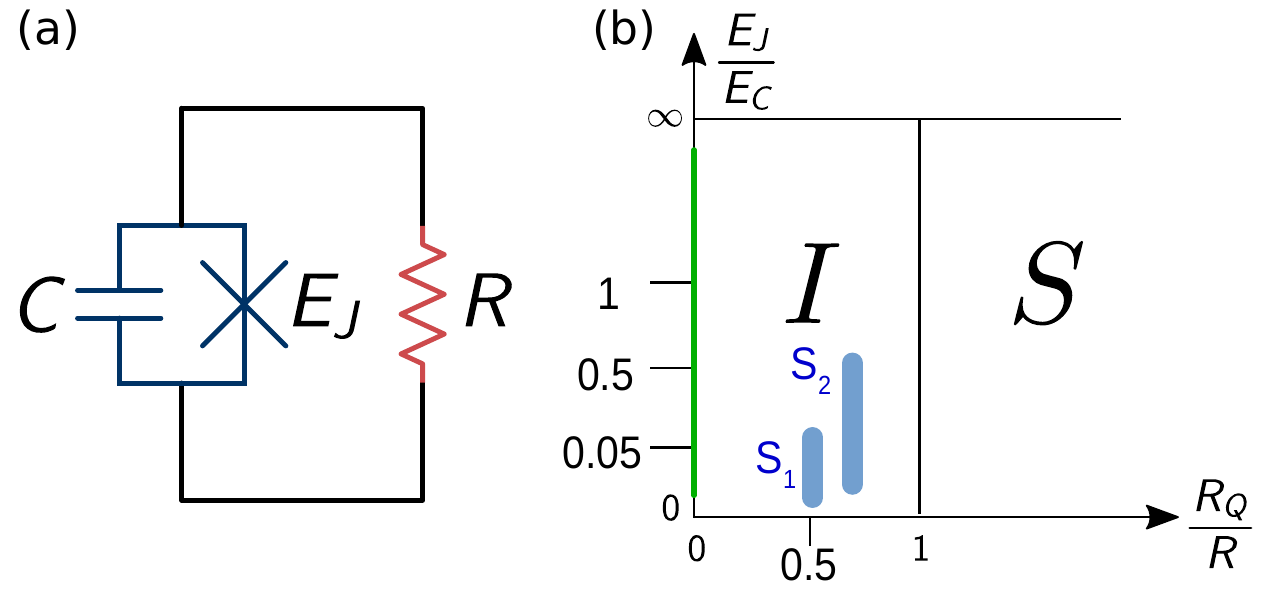}}
  \caption{\label{fig1}(a) A Josephson junction connected to a resistor $R$
  (abbreviated as JJ+$R$). The junction's capacitance $C$ determines the
  charging energy $E_C = (2 e)^2 / 2 C$, while the transparency of the tunnel
  barrier and the superconducting gap set its Josephson coupling energy $E_J$.
  (b) Sketch of the Schmid-Bulgadaev phase diagram for the circuit in (a). In
  the phase $I$ ($S$), the junction is predicted to be insulating
  (superconducting) at zero temperature. The insulating phase is paradoxical
  because the left axis (green line, where $R = \infty$) is the location of
  the Cooper pair box family of superconducting qubits for which it is well
  known that the junction is superconducting. Similarly, our samples $S_1$ and
  $S_2$ are found to remain superconducting when lowering the temperature,
  even though they are supposed to be well inside the insulating phase.}
\end{figure}

In 1983, Schmid {\cite{schmid_diffusion_1983}} predicted that a
dissipation-driven quantum phase transition (DQPT) should occur for any
Josephson junction (JJ) connected to a resistance $R$ : When $R > R_Q = h / 4
e^2 \simeq 6.5 \text{k} \Omega$ the junction should be insulating at zero
temperature, while if $R < R_Q$, the junction should be superconducting (see
Fig. \ref{fig1}). The prediction was made more precise shortly after by
Bulgadaev {\cite{bulgadaev_phase_1984}}, and since then, many theoretical
works using different techniques
{\cite{guinea_diffusion_1985,aslangul_quantum_1985,zaikin_dynamics_1987,schon_quantum_1990,ingold_effect_1999,herrero_superconductor-insulator_2002,kimura_temperature_2004,kohler_quantum_2004,werner_efficient_2005,lukyanov_resistively_2007}}
have further confirmed it. Attempts to investigate this prediction
experimentally are scarce
{\cite{yagi_phase_1997,penttila_superconductor-insulator_1999,kuzmin_coulomb_1991}}
and these early experiments were all affected by technical limitations (see
Appendix \ref{former_exps}) that made their interpretation debatable. In this
work, we revisit this prediction using well-controlled linear response
measurements on the insulating side of the phase diagram, and we find no sign
of the junctions becoming insulating. By revisiting the theory, we provide
arguments explaining why, actually, no superconducting-to-insulating
transition is expected and we propose an alternative comprehensive physical
picture for this system.

Let us first motivate our work by explaining why the predicted phase diagram
is problematic. The left axis in the Schmid-Bulgadaev (SB) phase diagram [Fig.
\ref{fig1}(b)] corresponds to $R \rightarrow \infty$, where we can simply
remove the resistor from the circuit. In this limit, we are left with only the
junction represented as a pure Josephson element in parallel with the
junction's geometric capacitor $C$ defining the charging energy $E_C = (2 e)^2
/ 2 C$. Such a disconnected junction is known as a Cooper pair box (CPB) in
the domain of quantum circuits; it behaves as a nonlinear oscillator and has
been extensively investigated theoretically and experimentally
{\cite{bouchiat_quantum_1998-1,nakamura_coherent_1999,koch_charge-insensitive_2007,schreier_suppressing_2008}}.
In particular, for any junction with a nonzero Josephson coupling $E_J$, it
has finite charge fluctuations through the junction, in contradiction with it
being on the insulating side of the phase transition, and it was shown that
one can indeed drive finite ac supercurrents through the junction
{\cite{nguyen_current_2007-1}}. \ Furthermore, since the anharmonicity of the
CPB vanishes upon increasing the ratio $E_J / E_C$
{\cite{koch_charge-insensitive_2007}}, one expects (at least in the large $E_J
/ E_C$ range) the effect of a finite parallel resistance $R$ on this
nonharmonic oscillator to be similar to that on a harmonic oscillator
{\cite{grabert_quantum_1984,vool_introduction_2017}}: When $R$ is varied, the
phase and charge fluctuations have no abrupt change at $R = R_Q$. Approaches
that go beyond considering the junction as a pure inductor
{\cite{joyez_self-consistent_2013,nigg_black-box_2012}} confirm this intuition
down to the moderately large $E_J / E_C$ range : They predict a
superconductive junction that smoothly retrieves the ``bare'' (with no
resistor) CPB behavior as the environment impedance gets large and cold. More
generally, any Josephson junction connected to a large impedance $Z$ is
intuitively expected to smoothly recover the (superconducting) behavior of the
CPB in the $Z \rightarrow \infty$ limit. This was confirmed theoretically in
the specific case of a purely inductive environment in Ref.
{\cite{koch_charging_2009}}. In summary, several known theoretical results
{\cite{koch_charge-insensitive_2007,grabert_quantum_1984,vool_introduction_2017,joyez_self-consistent_2013,nigg_black-box_2012,koch_charging_2009}},
many experimental results
{\cite{bouchiat_quantum_1998-1,nakamura_coherent_1999,koch_charge-insensitive_2007,nguyen_current_2007-1}}
and intuitive expectations in simple limits are consistent among themselves
and conflict with the prediction of the insulating phase shown in Fig.
\ref{fig1}(b).\begin{figure}[b]
  \resizebox{8.5cm}{!}{\includegraphics{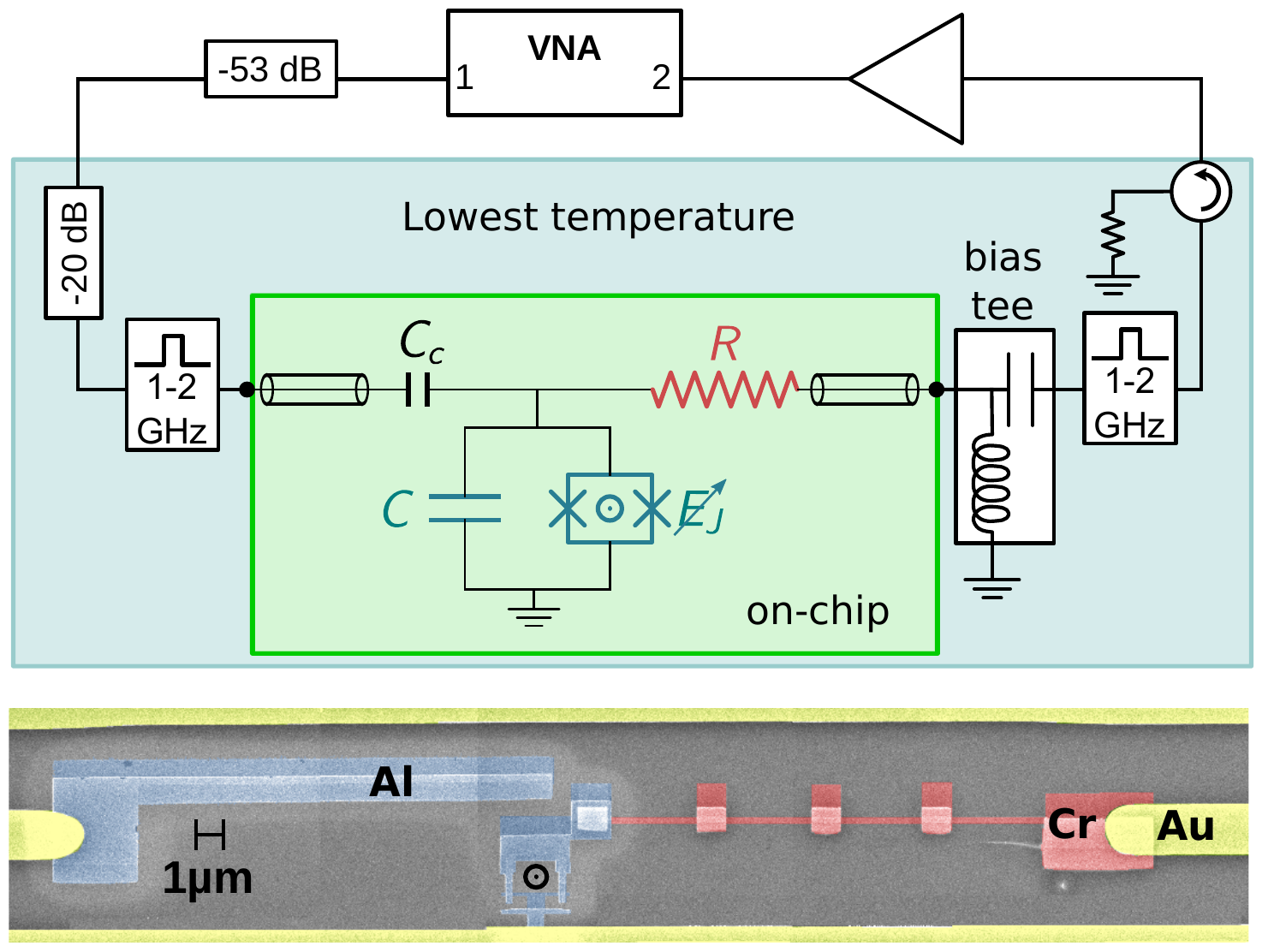}}
  \caption{\label{Exp_Setup} Top: Simplified schematics of the experimental
  setup. Bottom: One of the samples measured. Two SEM micrographs are stitched
  to show the entire central part and colorized to evidence the different
  metals used (see Appendix \ref{fabrication} for fabrication details).}
\end{figure}

\section{Experiment}

In order to test the SB prediction, we designed an experiment that closely
implements the circuit of Fig. \ref{fig1}(a) while allowing us to probe the
linear response of Josephson junctions in ac. A schematics of the experiment
and a micrograph of a sample are shown in Fig. \ref{Exp_Setup} and the main
sample parameters are given in Table \ref{sample_parameters}. Instead of a
single junction, we use a superconducting quantum interference device (SQUID)
behaving as an effective tunable Josephson junction : By applying a magnetic
flux $\Phi$ in the SQUID loop, its Josephson coupling energy is tuned as $E_J
\simeq E_{J \max}  | \cos (\pi \Phi / \Phi_0) |$ with $\Phi_0 = h / 2 e$ the
flux quantum. The input capacitor $C_c$ is chosen small enough that, at the
measurement frequency, it essentially converts the input ac signal into a
current source for the parallel junction-capacitance-resistance system. This
current is split between these components according to their admittance. The
fraction of the current flowing through the resistor is routed off chip to a
microwave bias tee. The dc port of the bias tee is shorted to ground, closing
the circuit in dc and ensuring there is no dc bias applied on the junction. At
the high-frequency port of the bias tee, the ac signal coming from the
resistor is sent through circulators and filters to a chain of microwave
amplifiers with an overall gain $\tmop{of}$ 106 dB. We used microwave
simulations of the circuit to check that in this design, the actual impedance
seen by the junction is close to $R \parallel C$ up to frequencies well above
$(R C)^{- 1}$ (note that the impedance to ground of the circuit following the
resistor is negligible compared to $R$ at all frequencies). We use a vector
network analyzer to perform continuous-wave homodyne measurements of the
transmission $S_{21}$ through the sample. Although in this setup we measure
variations of the fraction of the ac current flowing through the resistor,
they are directly related to the variations of the junction admittance (see
Appendix \ref{explainS21}).

\begin{center}
  \begin{table}[t]
    \begin{tabular}{|c|c|c|c|c|}
      \hline
      Sample & $E_c / k_B  \left( \text{K} \right)$ & $E_{J \max} / k_B 
      \left( \text{K} \right)$ & $R \left( \text{k} \Omega \right)$ & $C_c 
      (\tmop{fF})$\\
      \hline
      1 & 2.6 & 0.12 & $12$ & 0.3\\
      \hline
      2 & 0.64 & 0.39 & $8$ & 0.3\\
      \hline
    \end{tabular}
    \caption{\label{sample_parameters}Main sample parameters. See Appendix
    \ref{fabrication} for details on their determination.}
  \end{table}
\end{center}

\begin{figure}[tb]
  \resizebox{8.5cm}{!}{\includegraphics{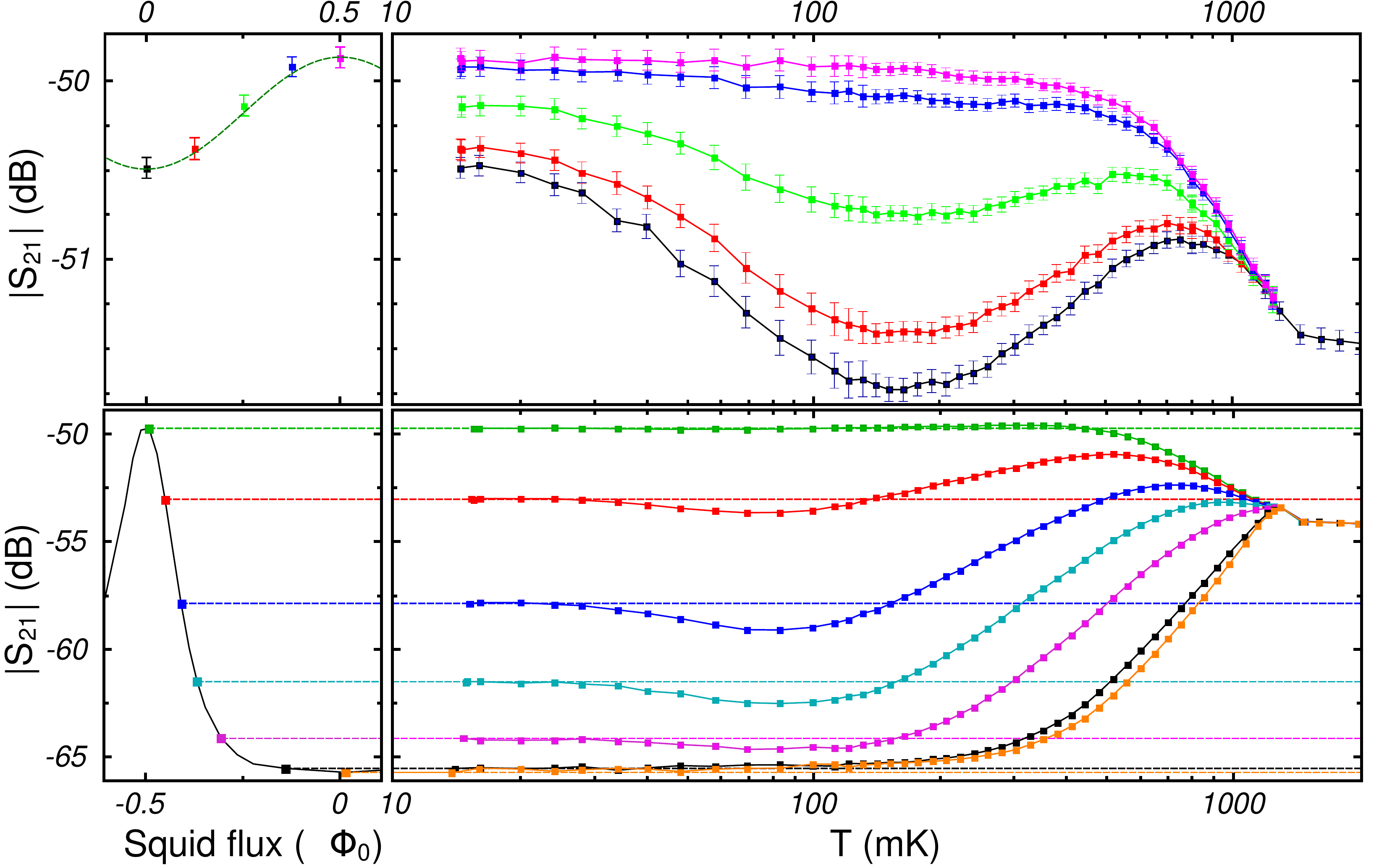}}
  \caption{\label{exp_data} Measured modulus of the transmission $S_{21}$ (as
  a power ratio) for sample 1 (top panels) and 2 (bottom panels). Left panels:
  $| S_{21} |$ as a function of the flux through the SQUIDs at the base
  temperature. The modulation is periodic with the flux (data not shown), as
  usual for a SQUID; only half a period is represented. Note that the position
  of the zero flux is different in the top and bottom panels. Right panels: $|
  S_{21} |$ for several flux values (using the same colors as on the left
  panels) as a function of the temperature. For sample 2, the error bars are
  smaller than the symbols used (note the larger vertical scale). }
\end{figure}

The operating conditions of the experiment are subject to constraints that we
now detail. First, in order to improve our sensitivity to the junction's
admittance {\cite{note1}}, the measurements need to be performed at a
frequency well below the ``plasma frequency'' $\omega_p = (C
L_J^{\tmop{eff}})^{- 1 / 2}$ of the junction so that, as seen from the input
capacitance, the ac current through $C$ is negligible. The current is then
essentially divided between the resistor and the junction's effective
inductance $L_J^{\tmop{eff}}$, should it exist, in proportion of their
respective admittance $1 / R$ and $1 / i L_J^{\tmop{eff}} \omega$. We selected
an operating frequency of order 1~GHz in order to simultaneously fulfill this
constraint (except in the vicinity of the maximal frustration of the SQUID)
and have a reasonably good noise temperature for our microwave amplifier.
Second, since we aim to probe the linear response of the junction at
equilibrium, the ac phase excursion must be $\delta \varphi \ll 2 \pi$, so
that the junction is properly described by an admittance $1 / i
L_J^{\tmop{eff}} \omega$. Assuming the worst case where all the current flows
through the resistor, this inequality restricts the ac amplitude at the sample
input $V_{\tmop{in}} \ll \Phi_0 / R C_c$ (that is, $P_{\tmop{in}} \ll - 50
\tmop{dBm}$ for the values used in the experiment; see below).
Correspondingly, all the measurements that we show here are taken in the low
power limit where $S_{21}$ no longer depends on the input power (see Appendix
\ref{linearity}). The last constraint also restricts the admissible input
power : The Joule power dissipated by ac current flowing through the resistor
should not raise its temperature significantly. We use the results of Ref.
{\cite{huard_electron_2007}} to estimate the electronic heating. Neglecting
electron-phonon cooling in the resistor, for the maximum $S_{21}$ value of $-
50 \tmop{dB}$, and at the input power of $- 70 \tmop{dBm}$ used for the sample
2 data at the lowest temperature ($T_{\tmop{ph}} = 13 \tmop{mK}$) in Fig.
\ref{exp_data}, one predicts an upper bound for the electronic temperature
rise of approximately $1.0 \tmop{mK}$ (0.5~mK for sample 1) close to the
junction (see Appendix \ref{heating}). Note that for such a low power level,
the signal-to-noise ratio at the input of the first cryogenic HEMT amplifier
is such that each data point necessitates averaging for about 20 min. Above
about 50 mK, electron-phonon cooling becomes effective (see Appendix
\ref{heating}); it is then possible to speed up the measurement by increasing
the excitation amplitude (still remaining in the linear regime) without
raising the electronic temperature.

In Fig. \ref{exp_data} we show the transmission $S_{21}$ for the two samples
we measured, for different flux through the SQUID and different temperatures.
On the left panels, we show $S_{21}$ as a function of the flux in the SQUIDs
at the lowest temperature. We observe that when the flux is zero in the SQUID,
the junction has the highest admittance ($S_{21}$ minimum), whereas its
admittance is minimum when the SQUID is frustrated with half a flux quantum in
the loop. On the right panels, we show the temperature dependence of $S_{21}$
for several values of the flux in the SQUIDs. We observe that in the low
temperature range, for any fixed value of the flux, $S_{21}$ reaches plateaus
indicating that the junction admittance saturates to a finite value. In other
words, at low temperature the modulation of $S_{21}$ with the flux proves that
the SQUID still carries supercurrent, and it shows no tendency to become
insulating at lower temperatures.

\section{Discussion }

Would the predicted insulating phase exist, the junctions would be in the
quantum critical regime where one expects the junction admittance to follow a
power law of the temperature {\cite{vojta_quantum_2003}}. This is clearly not
the case in our experiments. In a totally independent experiment with a
different objective, Grimm {\tmem{et al.}} {\cite{grimm_bright_2019}} have
recently observed that a SQUID with $E_J^{\max} / E_C \simeq 0.3$ in series
with a 32-k$\Omega$ resistance $(R_Q / R \simeq 0.2)$ had a clear dc
supercurrent branch that was modulated with the flux. We consider their
observation to support our results.

Together with the known $R \rightarrow \infty$ limit of qubits and the
observed superconducting junctions at $E_J / E_C \gtrsim 7$ and $R_Q / R \sim
0.6$ in Refs.
{\cite{penttila_superconductor-insulator_1999,penttila_experiments_2001}} (see
Appendix \ref{former_exps}), we conclude that the experimental observations
are consistent with a complete absence of the predicted insulating phase.

We now turn to theoretical considerations. In the first step, we revisit the
framework in which the SB prediction of a superconducting-to-insulating phase
transition was made. In the second step, we explain the exact nature of the
predicted transition and provide arguments according to which JJs are actually
not expected to become insulating in any Ohmic environment.

The SB prediction was cast using the model introduced by Caldeira and Leggett
(CL) {\cite{caldeira_quantum_1983}}, which describes a Josephson junction and
its capacitor (forming a CPB) analogous to a massive particle in a washboard
potential, coupling the particle position (the junction phase) to a bath of
harmonic oscillators that provide viscous damping. The corresponding
Hamiltonian is
\[ H = E_C N^2 - E_J \cos \varphi + \sum_n 4 e^2 \frac{N_n^2}{2 C_n} +
   \frac{\hbar^2}{4 e^2} \frac{(\varphi_n - \varphi)^2}{2 L_n}, \]
where $\varphi$ (resp. $N$) denotes the junction's phase (resp. number of
Cooper pairs on the junction capacitance) which are conjugate $[\varphi, N] =
i$, and the $\varphi_n$ (resp. $N_n$) denote the phase (resp. dimensionless
charge) of the harmonic oscillators. $H$ is not invariant upon $\varphi
\rightarrow \varphi + 2 \pi$, so that values of $\varphi$ differing by $2 \pi$
are naturally regarded as distinguishable states of the junction and $\varphi$
is said to be an ``extended phase.'' Correspondingly, $N$ has its spectrum in
$\mathbb{R}$, and we call it an {\tmem{extended}} charge too.

A unitary transformation $H' = U^{\dag} H U$ with $U = \exp (i \varphi N_R)$
and $N_R = \sum_n N_n$ (the charge passed through the resistor), yields
another Hamiltonian of interest
\[ H' = E_C (N - N_R)^2 - E_J \cos \varphi + \sum_n 4 e^2 \frac{N_n^2}{2 C_n}
   + \frac{\hbar^2}{4 e^2} \frac{\varphi_n^2}{2 L_n}, \]
where the CPB now couples to the environment through $N$, here representing
the number of transmitted Cooper pairs through the junction. Unlike $H$, $H'$
is evidently invariant upon the discrete translation $\varphi \rightarrow
\varphi + 2 \pi$ \ so that the values of $\varphi$ differing by $2 \pi$ can be
regarded as indistinguishable (wave functions in $\varphi$ are $2 \pi$
periodic), and the usual terminology is that $\varphi$ is a ``compact phase.''
In principle, $\varphi$ can still be described as an extended variable, in
which case the periodicity of the potential implies that wave functions in
$\varphi$ are Bloch functions $\Psi_q (\varphi) = \sum_n a_n (q) e^{i (n + q)
\varphi}$. Here, the ``quasicharge'' $q$ is a conserved quantity fixed by
initial conditions. However, a Bloch function with quasicharge $q$ can be
transformed to any other Bloch function by a global shift of the bath charge,
and the resistance is translationally invariant in both charge and phase (this
invariance being respected in the CL model {\cite{caldeira_quantum_1983}}).
Thus, in this resistively shunted JJ, states with different quasicharges can
be considered degenerate in the sense that no measurement on the circuit can
distinguish them after the initial charge shift of the bath has decayed
{\cite{note3}}. Thus one can choose to use only compact phase states $(q = 0
\tmop{mod} 2 \pi)$ for convenience. In this case, \tmtextit{N }has a discrete
spectrum in $\mathbb{Z}$ (even though there is no island in the circuit) and
the Josephson coupling term can be written as $E_J \cos \varphi = \frac{1}{2}
E_J \left( \sum_{N \in \mathbb{Z}} | N \rangle \langle N + 1 | + \text{H.c.}
\right)$ as customary for CPBs which we expect to recover in the $R
\rightarrow \infty$ limit.

With these provisos, $H$ and $H'$ operate on wave functions with different
symmetries; they almost seem to describe different physical systems. This
issue was known from the start and several theory papers considered the
suitability of either phase description for the system considered here, but no
clear-cut answer emerged (for an overview see Ref.
{\cite{mullen_resonant_1993}}). However, a unitary transformation cannot break
a symmetry of the system, and the contradiction resolves when one properly
transforms the boundary and initial conditions together with the Hamiltonian
{\cite{loss_effect_1991-1,mullen_resonant_1993}}. Provided this transformation
is carried out properly and barring any spontaneous symmetry breaking, $H$ and
$H'$ can be used indifferently to describe the system and any valid state of
the system should thus be representable with either $H$ or $H'$. As we mention
above, in the $R = \infty$ limit of the bare CPB, the phase is known to be
compact; hence, by continuity, compact states are also the states to consider
at finite $R$, unless one shows a spontaneous symmetry breaking of the
discrete phase translation invariance occurs, a phenomenon also known as the
``decompactification''
{\cite{schon_quantum_1990,apenko_environment-induced_1989}} of the phase (and
which goes along with an ``undiscretization'' of the charge).

\begin{figure}[t]
  \resizebox{8.5cm}{!}{\includegraphics{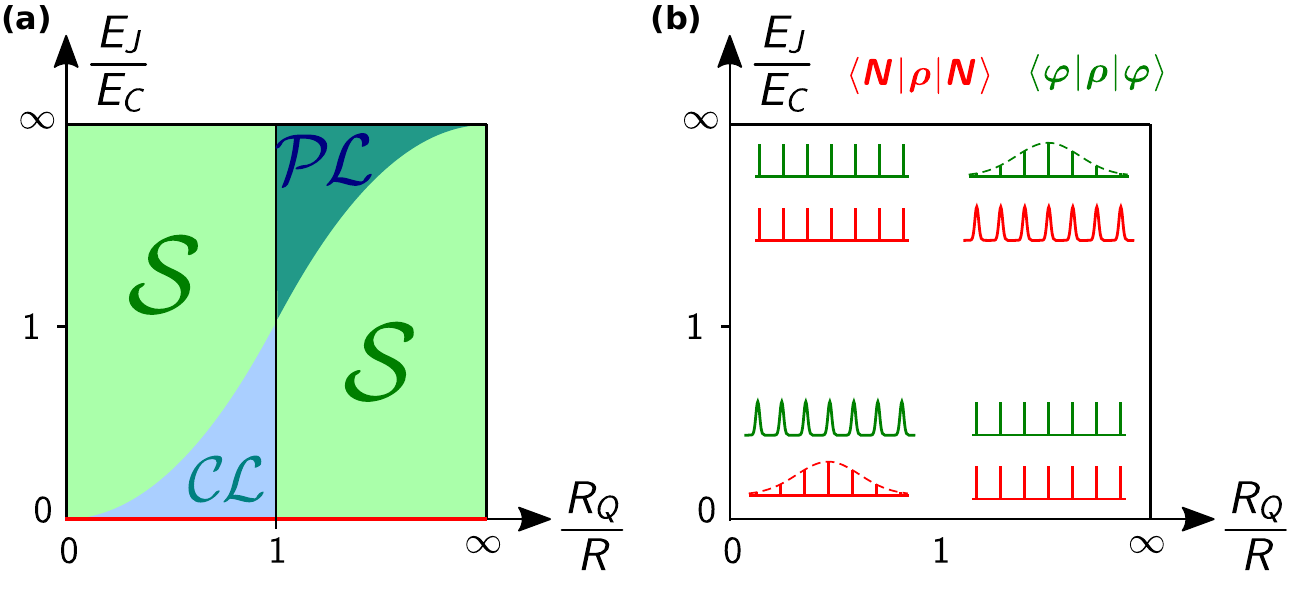}}
  \caption{\label{new_diagrams}(a) Reinterpreted Schmid-Bulgadaev phase
  diagram, in which the junction is superconducting everywhere, except for
  $E_J = 0$. In the $\mathcal{S}$ parts, the junction is superconducting with
  a fully delocalized charge and, correspondingly, a fluctuationless
  (classical) compact phase. Partial charge (phase) localization occurs in the
  $\mathcal{C}\mathcal{L}$ ($\mathcal{P}\mathcal{L}$) part. The classical
  phases $\mathcal{S}$ are artifacts which disappear when improving the model
  (see text and Appendix \ref{tightbinding}). (b) Final description of the
  junction's behavior in the parameter space. Drawings are sketches of the
  diagonal elements of the junction's reduced density matrix in an extended
  description (red, charge representation; green, phase representation). Close
  to the left (right) half of the upper (lower) axis, they nearly take the
  form of Dirac combs where the phase is almost a classical variable. In the
  lower left (upper right) sector, partial charge (phase) localization occurs,
  as in (a). The density matrix evolves continuously, interpolating between
  these limits, without any phase transition.}
\end{figure}

The SB theory is precisely all about dissipation causing spontaneous symmetry
breaking; we now describe the core ideas of this theory. Close to the bottom
axis of the phase diagram, in the so-called scaling limit where $E_C
\rightarrow \infty$ (which constrains $N = N_R$), $H'$ becomes equivalent to
the tight-binding model used in Refs
{\cite{guinea_diffusion_1985,aslangul_quantum_1985}} (see Appendix
\ref{tightbinding}). In this model, at low friction (low $R$), the junction's
zero-temperature reduced density matrix $\rho$ is completely delocalized in
the discrete charge basis, and thus corresponds to a perfectly localized
compact phase. For such a state, using an extended description for both charge
and phase, the diagonal of $\rho$ is a Dirac comb in both charge and phase
representation [Fig. \ref{new_diagrams}(b), bottom right]. For $R > R_Q$,
however, the discrete translational invariance symmetry of the charge is
broken and the charge localizes at a given value of $\langle N \rangle =
\tmop{Tr} \rho N$. In $\rho$, the result of this charge localization can be
seen as multiplying the charge Dirac comb by a bell-shaped function $b$ and
broadening each peak of the phase Dirac comb by convolving it with the Fourier
transform of $b$ [Fig. \ref{new_diagrams}(b), bottom left]. Across the
transition, the charge fluctuations (the width of $b$) vary continuously
{\cite{aslangul_quantum_1985}}, but the dc charge mobility $\mu$ (related to
the charge fluctuations according to the standard Green-Kubo relations; see
Appendix \ref{yvsmu}) is predicted to vanish (resp. diverge) for $R > R_Q$
(resp. $R < R_Q$) at $T = 0$, hence, the prediction of a
superconducting-to-insulating transition. \

At the time of the prediction, this mobility argument was often associated
with the simple picture of infinite polaronic trapping in the insulating phase
$R > R_Q$ and the corresponding suppression of the coherence $E_J \langle \cos
\varphi \rangle$ between charge states. More elaborate renormalization group
(RG) flow arguments
{\cite{schmid_diffusion_1983,bulgadaev_phase_1984,guinea_diffusion_1985,schon_quantum_1990,lukyanov_resistively_2007,fisher_quantum_1985}}
led to the conclusion that an immobilization of the junction charge indeed
occurred in the whole domain where the cutoff frequency of the Ohmic damping
is the fastest dynamics in the system, i.e., $E_J / E_C < (R_Q / R)^2$ [see
part $\mathcal{C}\mathcal{L}$ in Fig. \ref{new_diagrams}(a); note that our
experimental parameters are in this zone].

However, works on the closely related spin-boson Problem (SBP; the CL model
is an infinite-spin generalization of the SBP) have shown that the picture of
infinite polaronic trapping is too naive. In this system, the spin and the
bath entangle in the ground state, involving an infinite number of bosonic
excitations {\tmem{and}} yielding resilient finite coherences (possibly very
small)
{\cite{spohn_ground_1989,hur_entanglement_2008,bera_stabilizing_2014,bera_generalized_2014}}
that depend algebraically on the UV cutoff of the Ohmic bath. In the CL model
itself, perturbation theory in $E_J$ shows as well that, while $E_J \langle
\cos \varphi \rangle = 0$ at zeroth order, $E_J \langle \cos \varphi \rangle
=\mathcal{O} (E_J^2 / E_C )$ (as in the bare CPB) at the next order
{\cite{grabert_unpublished_nodate}} for any $R > R_Q / 2$. Hence, it is no
longer believed that the coherences vanish in the ``insulating phase,'' and
this has dramatic consequences: (i) It enables a finite supercurrent flow (as
evidenced by our experiments), and (ii) previously calculated dc charge
mobility does not describe the actual transport properties, because it does
not take into account the inductive behavior associated with the supercurrent
(see Appendix \ref{yvsmu}). The qualitative explanation for the robustness of
the coherence is that the inductive response of the junction shunts the
low-frequency modes of the environment that were supposed to fully suppress
the coherence {\cite{joyez_self-consistent_2013}}. In this new understanding
of the (previously believed) insulating phase, the partially localized charge
states are similar to those of the bare CPB, and they very naturally coincide
with them in the $R \rightarrow \infty$ limit. The difference between the
resistively shunted junction and the CPB with an island is that in the first
case there is a degenerate continuum of localized charge states at all values
of $\langle N \rangle$, while in the second case where no dc current can flow,
$\langle N \rangle$ is pinned and the ground state is unique.

Close to the top axis of the phase diagram, one follows similar reasoning in
the ``dual'' picture {\cite{weiss_quantum_2012}}, where charge and phase are
interchanged. One then starts from a tight-binding description of Wannier
states for the phase located in the different wells of the cosine potential
(and where the strength of the friction is inverted
{\cite{weiss_quantum_2012}}). Mirroring what occurs on the bottom axis, this
duality predicts that the diagonal of $\rho$ is again a Dirac comb in both
charge and phase representations [upper left of Fig. \ref{new_diagrams}(b)] at
low friction (large $R$) and that a smooth spontaneous symmetry breaking
transition to partial ``phase localization'' occurs for $R < R_Q$ [part
$\mathcal{P}\mathcal{L}$ in Fig. \ref{new_diagrams}(a)]. We thus identify this
transition as a {\tmem{progressive}} decompactification of $\varphi$. This
shows that a generic decompactified phase state is the dual of a CPB state,
i.e., a superposition of classical phase states differing by $2 \pi$ in
several adjacent wells of the cosine. To our knowledge, this is the first time
the decompactification process is clarified and it is a key result as it shows
this spontaneous symmetry breaking does {\tmem{not}} yield generic extended
phase states, contrary to what was generally assumed so far (see Appendix
\ref{compact-extended}). In particular, Schmid and subsequent authors treated
$\varphi$ as extended, which led them to attribute an insulating character to
the ``delocalized phase'' in all the wells of the cosine (for $R > R_Q$).
However, when considering a compact phase, the junction is insulating only
when the phase is completely delocalized \tmtextit{within} one period (all
coherences vanishing: $\langle \cos n \varphi \rangle = 0$, \ $\forall n \in
\mathbb{N}^{\ast}$), meaning that the diagonal of $\rho$ is completely flat in
the phase representation.

In Fig. \ref{new_diagrams}(a), we show our reinterpretation of the SB phase
diagram, where the junction is superconducting everywhere, except at $E_J /
E_C = 0$, in the so-called {\tmem{scaling limit}}. Note that in actual
implementations, $E_C$ is always finite, so that the insulating state of the
scaling limit can be achieved only by choosing $E_J = 0$, i.e., trivially, an
already fully insulating junction (even in the normal state). This
reinterpreted diagram is in agreement with experiments and resolves the
conflicts mentioned in the Introduction. At this point, the vertical
boundaries at $R = R_Q$ which remain from the SB prediction are continuous
transitions from fluctuationless phase states to states having finite
zero-point phase fluctuations, i.e., classical-to-quantum transitions.
However, one can show these transitions arise from properties of the uncoupled
bosonic bath (Ref. {\cite{guinea_diffusion_1985}} and Appendix
\ref{tightbinding}), and one expects that a better treatment (taking into
account the aforementioned entanglement of the junction with the bath) should
restore finite phase fluctuations in the phases $S$, turning this transition
into a crossover.

The emergent understanding of this system is represented pictorially in Fig.
\ref{new_diagrams}(b): The junction is superconducting everywhere and its
reduced density matrix evolves continuously as a function of the parameters,
interpolating between the limit cases depicted. From this diagram one sees
that when the effective Josephson Hamiltonian is deemed adequate to model a
Josephson junction (see Appendix \ref{effectiveH}), the junction phase can be
essentially regarded as compact (and one can use the discrete charge basis of
a CPB) below the main antidiagonal, while one expects a partial
decompactification of the phase above that antidiagonal.

Obviously, generic extended phase states do not have the appropriate
symmetries within this understanding. Consequently, assuming an extended phase
to describe the low-energy states in such a system is at best approximate or
it appeals to (perhaps unspoken) ingredients external to the CL model. Yet,
many predictions (besides the DQPT) were made assuming an extended phase and
have been checked to well describe the Josephson physics. This raises the
question of when can one safely use such a description? A nonoperative answer
is that such a description is fine as long as interference effects that would
appear in a proper treatment of the phase (more or less complete) translation
invariance play no significant role.

Before closing this discussion, let us comment on the striking dips observed
in the temperature dependence of the transmitted power near $T \sim 100
\tmop{mK}$ corresponding to a maximum of the junction admittance. They can be
understood at a qualitative level using the usual charge description of the
CPB (consistent with the above discussion), assuming the resistance is large
enough. In the regime $E_J \ll E_C$ and at very low temperature, the state of
the CPB is nearly a classical state at the minimum of a charging energy
parabola with a given $N$. This state nevertheless has quantum fluctuations
that can be computed by second-order perturbation theory, with virtual
transitions through the neighboring charge states. This process results in an
effective Josephson coupling for the ground state $E_J^{\tmop{eff}} = E_J
\langle \cos \varphi \rangle \propto E_J^2 / E_C$, the energy denominator
$E_C$ being the energy of the virtual states. At finite temperatures $k_B T
\lesssim E_C$, low-energy modes of the resistance are thermally populated;
they can lend their energy to the virtual state, lowering the energy
denominator and thus increasing the effective Josephson coupling. \ At higher
temperatures, thermal fluctuations eventually reduce the gap of Al, reducing
the Josephson coupling.

\section{Meta-discussion:}

Given that the present work contradicts more than 35 years of literature on
the understanding of a Josephson junction in a resistive environment, one may
rightfully wonder if alternative explanations of our results could exist. One
can hypothesize that

\

\begin{quoteenv}
  (i) we might over- or misinterpret our experimental data and that of Grimm
  {\tmem{et al.}} {\cite{grimm_bright_2019}} when we conclude that the
  junctions remain superconducting at low temperatures, and\\

  (ii) there could be hidden flaws in our theoretical analysis of the CL
  model which leads us to conclude that no insulating phase is expected in
  JJ+$R$ systems,
\end{quoteenv}

such that the original SB prediction regarding JJs could stand. Within these
hypotheses, signatures of the insulating state could, for instance, appear
only out of the experimental windows for some reason to be worked out, making
our experimental data compatible with the original prediction. However, we
stress that the inconsistency of the insulating phase with the intuitive
limits that we point to in the Introduction would still need to be addressed.

We thus encourage experimental and theoretical work in this domain that could
complete, clarify or correct our findings, in the hope that the community soon
reaches consensus on the expected behavior of this key quantum component in
the presence of an environment.

\section{Conclusions}

Our experimental results show no evidence of the superconducting-to-insulating
DQPT in Josephson junctions predicted by Schmid and Bulgadaev, contrary to
present widespread expectations. We provide theoretical arguments according to
which the superconducting coherence in JJs is actually resilient to
dissipation, thereby barring the occurrence of that DQPT in JJ+$R$ systems
(the DQPT does occur in {\tmem{non-superconducting}} 1D systems, however; see
Appendix J). We reach a global and consistent qualitative description of JJs
with an environmental impedance that dovetails all well-known limits. As an
important by-product, our analysis for the first time clearly exposes how
phase decompactification occurs in Josephson junctions. This shows that
generic extended phase states are not rigorous solutions for this system,
hopefully settling decades of controversies. Our work also highlights that
there are presently no comprehensive and quantitative predictions for the
effect of dissipation on the CPB able to reproduce our results. Finally, our
results prompt for a critical reexamination of the works where the
Schmid-Bulgadaev prediction regarding Josephson junctions was used to draw
predictions for other systems such as superconducting nanowires proposed to
implement quantum phase slip junctions
{\cite{kerman_fluxcharge_2013,mooij_superconducting_2006,ulrich_dual_2016-1}}.

\section*{Acknowledgments}

The authors are grateful to M. Devoret, B. Dou{\c c}ot, S. Florens, M.
Hofheinz, C. Mora, I. Safi, H. Saleur, P. Simon and N. Roch for stimulating
discussions and suggestions. The technical assistance of Pascal S{\'e}nat and
SPEC's Nanofabrication lab is acknowledged, and we thank other members of the
Quantronics group for their constant support. This work is supported in part
by ANR Grants No. ANR-15-CE30-0021-01 and ANR-18-CE47-0014-01, the ANR-DFG
Grant JosephSCharli and by the LabEx PALM Project No. ANR-10-LABX-0039-PALM.
J. A. and J. S. are supported by the German Science Foundation under Grant
AN336/11-1 and the Center for Integrated Quantum Science and Technology
(IQST). C. A. aknoweldges funding funding from the European Research Council
under the European Union's Horizon 2020 program (ERC Grant Agreement No.
639039).

\

\appendix\section{Former experimental tests}\label{former_exps}

The SB prediction has been researched experimentally
{\cite{yagi_phase_1997,penttila_superconductor-insulator_1999,kuzmin_coulomb_1991,penttila_experiments_2001}},
but the scaling laws expected to be the hallmark of the predicted quantum
critical regime have not been thoroughly investigated.

In these experiments, the junction and its Ohmic shunt resistance $R$ were
typically ``current biased'' using a voltage source in series with a large
resistor $R_{\tmop{bias}}$>$R$ and measured using a lock-in technique at
frequencies $f_{\tmop{LI}} \sim 100 \tmop{Hz}$ or below. Could such a setup
properly measure the linear response of the junction?

For junctions with small critical current, it is well known that spurious
noise in the setup rapidly reduces the apparent maximum supercurrent
{\cite{joyez_josephson_1999-1,steinbach_direct_2001,vion_thermal_1996}}, and
particularly so for underdamped junctions, i.e., when $E_J / E_C \gg (R_Q /
R)^2$. However, even when the technical noises are completely eliminated, a
lock-in measurement has intrinsic limitations when the junction's admittance
becomes smaller than $1 / R$. In that case, keeping small phase excursion in
these setups requires an ac voltage excitation at the junction $V_{\tmop{ac}}
\ll \Phi_0 f_{\tmop{LI}} < 1 \tmop{pV}$ which, even taking into account the
resistive bridge division $R / (R + R_{\tmop{bias}})$, is several orders of
magnitude smaller than required to have a sufficient signal-to-noise ratio in
lock-in measurements. Thus, the former experiments aiming to test the DQPT
could not properly measure the linear response of junctions with very low
admittances: Several periods of the cosine were explored, rapidly averaging
any small supercurrent to zero. In contrast, in our setup, measuring at much
higher frequencies enables us to use larger excitation voltages while
remaining in the linear phase response regime, even when the admittance of the
junction becomes very low.

On the other hand, it is easy to observe the supercurrent branch of junctions
having a large critical current, even with an imperfect setup, because the
junction very effectively shunts noise. Indeed, the authors of Refs.
{\cite{penttila_superconductor-insulator_1999,penttila_experiments_2001}}
found that a superconducting branch was observed for all junctions supposed to
be in the insulating phase, provided that $E_J / E_C \gtrsim$7. At the time of
this result, the discrepancy with the DQPT prediction was resolved by arguing
that the observed superconducting state was a transient and that the true
equilibrium insulating state would be reached only after a possibly
cosmologically long time
{\cite{penttila_experiments_2001,schon_quantum_1990,penttila_superconductor-insulator_1999}}.
The argument given was that when the junction's (extended) phase starts
localized in one well of the cosine potential, it will eventually delocalize
in all other wells of the cosine by tunneling (and this delocalized state was
assumed insulating), but the tunneling rate becomes immeasurably small for
large $E_J / E_C$. However, when timescales become very long and energies very
small, one should seriously reconsider all other approximations made in the
modeling, such as, for instance, neglecting the level separation in the
electrodes. When considering a compact phase, such a slow phenomenon simply
does not exist : The phase is always instantly delocalized in all wells of the
cosine, and moreover, that state is superconducting. The superconducting state
observed in these experiments was then the genuine equilibrium state.

\section{Fabrication details}\label{fabrication}

The fabrication of the sample starts from a gold 50-$\Omega$ coplanar
waveguide (CPW) defined by optical lithography and providing the input and
output ports for the microwave signals. The central conductor of the
transmission line is interrupted on a length of 38 $\mu$m, creating a cavity
in which the resistor and junctions are fabricated in two subsequent steps,
using $e$-beam lithography and evaporation through suspended masks. The
resistor consists of a 8.5-nm-thick, approximately 100- nm-wide and
16-\text{$\mu$m}-long Cr wire, periodically overlapped with 45-nm-thick, $1
\times 1 \left( \mu \text{m} \right)^2$ Cr cooling pads. One end of the
resistor connects to the output transmission line. The junctions are produced
by standard double-angle evaporation of aluminum. The SQUID was connected on
one side to the ground plane of the CPW and on the other side, to the other
end of the Cr resistor. Microwave simulations of the circuit were used to
check that in this design, the actual impedance seen by the junction is close
to $R \parallel C$ up to frequencies well above $(R C)^{- 1}$. In order to
meet this condition, it is important that the whole SQUID + resistor layout is
very compact to avoid stray inductances and capacitances.

\subsection{Determination of the sample parameters}

Since the values of $E_J$ and $R$ cannot be independently measured directly on
the sample, the values reported in Table \ref{sample_parameters} come from the
room-temperature measurements of the resistance of several other junctions and
resistors having the same dimensions and fabricated at the same time on the
sample. From the scatter of these measurements, the values reported are
believed to be accurate within $\pm 15\%$. The value of $E_c$ is estimated
from the area of the junction using the commonly used value $100 ~ \tmop{fF}
(\mu \text{m})^{- 2}$ for the capacitance per unit area of aluminum-aluminum
oxide junctions. The value of the coupling capacitance is obtained from
microwave simulations.

\section{Joule heating in the resistor}\label{heating}

Here we show that for the measurements shown in Fig. \ref{new_diagrams}, the
Joule power dissipated in the chromium resistor does not substantially raise
the electronic temperature. For this, we rely on the analysis of heating in
diffusive wires detailed in Ref. {\cite{huard_electron_2007}}, where it is
assumed that the electron temperature can be well defined locally, i.e, that
the thermalization between electrons occurs faster than their diffusion
through the wire and that we can neglect the radiative cooling of the wire. In
this reference, the diffusive wire is supposed to be connected to two
normal-metal reservoirs at both ends, and these reservoirs are supposed to be
large enough so that their electronic temperature is equal to the phonon
temperature. In our case, on the junction side the Cr wire is connected to
superconducting Al which blocks any heat exchange at very low temperatures. We
can nevertheless obtain the electronic temperature at this point by
considering the results of Ref. {\cite{huard_electron_2007}} in the middle of
a wire with twice the length, twice the resistance and twice the dissipated
power.

We first evaluate the maximum Joule power $P_R$ dissipated in the Cr resistor
for the measurements performed at the lowest temperature ($13 \tmop{mK}$) in
Fig. \ref{new_diagrams}. This power is proportional to the power
$P_{\tmop{out}}$ at the output of the sample by
\[ P_R = \frac{R}{Z_0} P_{\tmop{out}}, \]
where $Z_0 = 50 \Omega$ is the impedance of the microwave circuitry and
\[ P_{\tmop{out}} = P_{\tmop{VNA}} 10^{\left( \frac{S_{21} - G_{}}{10} \right)
   }, \]
where $P_{\tmop{VNA}}$ is the power at the vector network analyzer (VNA)
output, $S_{21}$ is the measured transmission of the setup (in dB) and $G = +
106 \tmop{dB}$ the overall gain (in dB) of the microwave chain from the sample
output to the VNA input. For sample 2, using the maximum value $\tmop{Max} (|
S_{21} |) = - 50 \tmop{dB}$, $P_{\tmop{VNA}} = + 3 \tmop{dBm}$, and $R_{} = 8
\text{k} \Omega$, this leads to a maximum $P_R \simeq 80 \text{aW}$ \ [for
Sample 1: $\tmop{Max} (| S_{21} |) = - 50 \tmop{dB}$, $P_{\tmop{VNA}} = - 4
\tmop{dBm}$, and $R_{} = 12 \text{k} \Omega$ give a maximum $P_R \simeq 25
\text{aW}$].

Looking for an upper bound for the electronic temperature, we consider the
simple ``interacting hot-electron'' limit, where electron-phonon interaction
in the wire is neglected, so that cooling occurs only through diffusive
electronic exchange with the reservoir (here the gold central conductor of the
CPW). In this limit, the maximum temperature (reached in the middle of the
wire in Ref. {\cite{huard_electron_2007}} and at the Cr-Al interface in our
case) is
\[ T_{\max} = \sqrt{T_{\tmop{ph}}^2 + \frac{3}{4 \mathit{\pi}^2} \left(
   \frac{e}{k_B} \right)^2 2 R 2 P_R} \]
where $T_{\tmop{ph}}$ is the phonon temperature in the reservoir. At the
lowest temperature $T_{\tmop{ph}} = 13 \tmop{mK}$, and for the above values
this yields
\begin{eqnarray*}
  T_{\max} & \simeq & T_{\tmop{ph}} + \left\{\begin{array}{l}
    1 \tmop{mK} \text{for sample 2},\\
    0.5 \tmop{mK} \text{ for sample 1},
  \end{array}\right.
\end{eqnarray*}
which sets an upper bound for the electronic temperature of the
electromagnetic environment in our experiments. These considerations show that
in the entire experimental range, Joule heating of the resistor was
negligible.

In the above analysis, the thick intermediate pads incorporated in the wire
design (see Fig. \ref{Exp_Setup}) play absolutely no role. They are meant to
increase electron-phonon coupling, but they are effective only at higher
temperature as we now discuss. At the maximum power dissipated in the
resistor, we can estimate the electronic temperature $T_{\Sigma} = (P_R^2 /
\Sigma \Omega)^{1 / 5}$ {\cite{huard_electron_2007}} that would be reached if
only electron-phonon cooling was taking place. Taking the entire volume of the
resistive wire and of the intermediate cooling pads $\Omega \simeq 0.20 \left(
\mu \text{m} \right)^3$ and assuming the standard electron-phonon coupling
constant $\Sigma \simeq 2 \tmop{nW} \left( \mu \text{m} \right)^{- 3}
\text{K}^{- 5}$ gives $T_{\Sigma} \simeq 36 \tmop{mK}$ (for sample 2). We
could thus increase measurement power at temperatures above $50 \tmop{mK}$ in
order to speed up the measurements while still not heating the electrons.

\

\section{Checking the linearity of the response}\label{linearity}

\begin{figure}[tb]
  \resizebox{8.5cm}{!}{\includegraphics{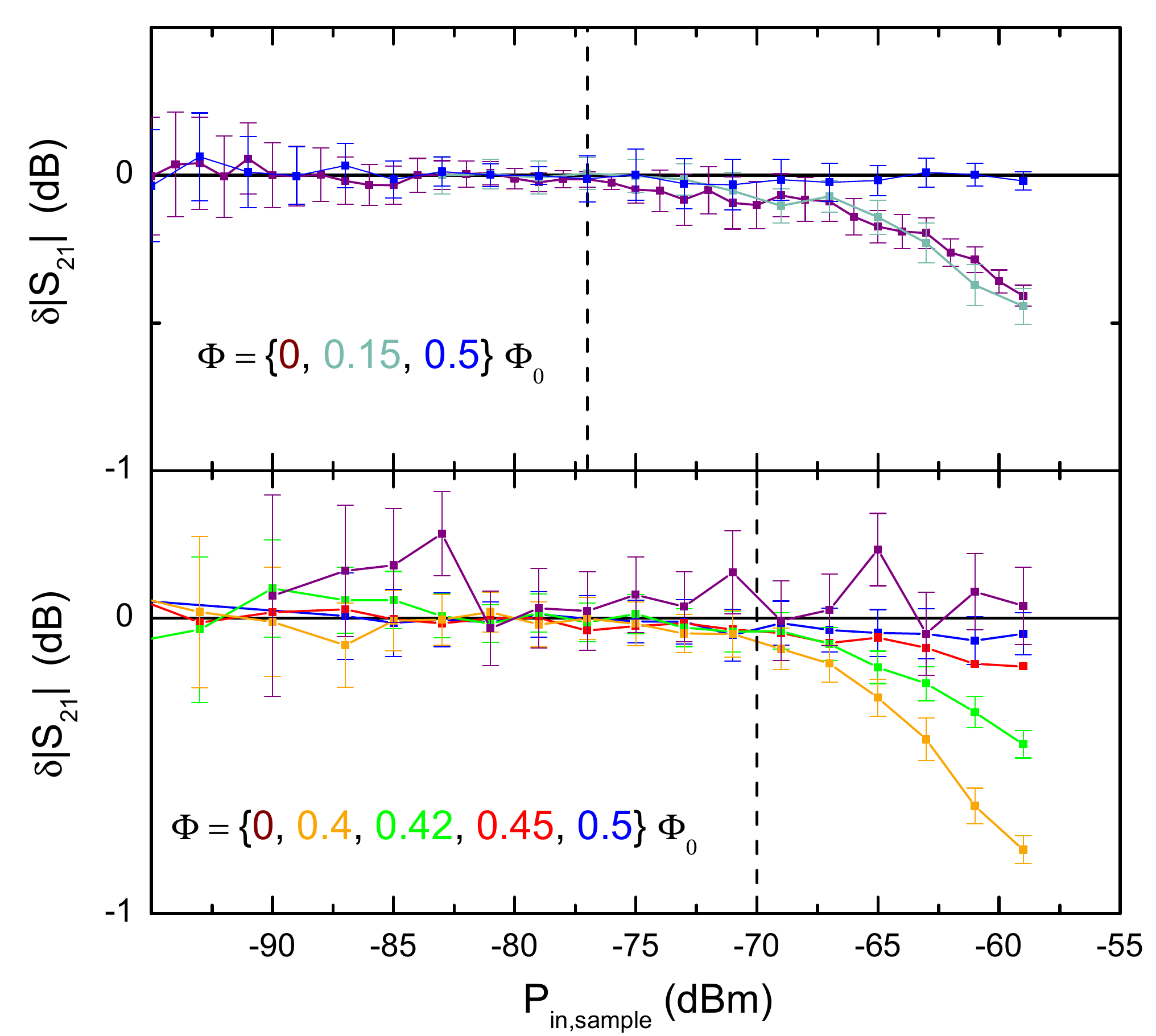}}
  \caption{\label{linearity_check} Variations of the transmission through the
  samples (top, sample 1; bottom, sample 2), as a function of the power at the
  sample input, for different values of the flux through the SQUIDs, at the
  lowest temperature (variations are taken with respect to the value at
  $P_{\tmop{in}} = - 80 \tmop{dBm}$). The size of the error bars does not vary
  monotonically because the averaging time is increased when reducing the
  power. The dashed lines indicate the power levels that were chosen to take
  the data shown in Fig. \ref{exp_data}.}
\end{figure}In order to ascertain that we properly measure the linear response
of the junction, we check that $S_{21}$ no longer depends on applied power at
low power. In Fig. \ref{linearity_check} we show the variations of $| S_{21}
|$ as a function of the applied measurement power for various fluxes in the
two samples at the lowest temperature (13 mK). We indeed observe that in the
low-power range $| S_{21} |$ no longer changes, confirming that we measure the
linear response and that we are not heating the resistor. We used such
measurements to choose the operating power for the data presented in Fig.
\ref{exp_data}, selecting the value at the end of the horizontal plateau
(shown as the dashed vertical line in Fig. \ref{linearity_check}), i.e., $- 77
\tmop{dBm}$ for sample 1 and $- 70 \tmop{dBm}$ for sample 2.

\section{Link between the measured $S_{21}$ and the Josephson-junction
parameters}\label{explainS21}

\begin{figure}[tb]
  \resizebox{5cm}{!}{\includegraphics{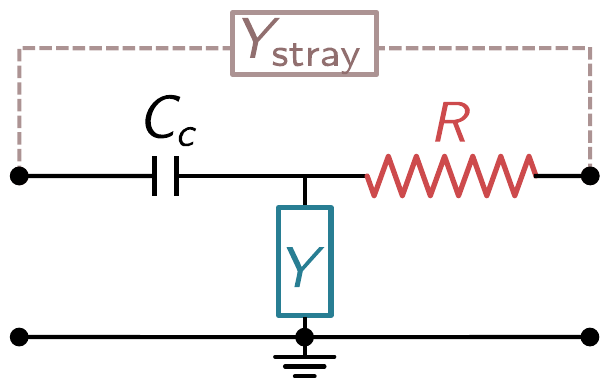}}
  \caption{ \label{quadrupole_model}Quadrupole model of the on-chip components
  for the calculation of the transmission $S_{21}$. In the ideal case where
  the external circuit can be fully calibrated by measuring reference samples,
  $S_{21}$ would depend only on $C_c$, $R$, and $Y$, the admittance of the
  junction. In our experiment this full calibration is not performed, and a
  weak stray admittance $| Y_{\tmop{stray}} | \ll | i C_c \omega |$ very
  likely dominates our measurements at lower values of $S_{21}$, when the
  junction admittance is large $(| Y | R \gg 1)$.}
\end{figure}

Assuming a probe signal has a low enough amplitude, the circuit between the
input port and the output port of the sample chip is then a linear quadrupole
as depicted in Fig. \ref{quadrupole_model}, with the junction described as a
linear admittance $Y (\omega)$ corresponding to the parallel combination of
the capacitance and the SQUID of Fig. \ref{Exp_Setup}. For such a quadrupole,
the input and output waves' amplitudes at both ports (assumed to have the
standard microwave characteristic impedance $Z_0 = 50 \Omega$) are related by
an $S$ matrix {\cite{pozar_microwave_2011}}. Considering that at the frequency
of the experiment $(i C_c \omega)^{- 1} \gg R \gg Z_0$, the $S$ matrix of the
sample is approximately
\[ S \simeq I + S_{21} \sigma_x \]
where $I$ is the $2 \times 2$ identity matrix, $\sigma_x$ the Pauli matrix,
and \
\[ S_{21} = \frac{2 i C_c \omega Z_0}{1 + R Y (\omega)} \]
is the transmission amplitude from the input to the output port.

Thus, in principle a measurement of the (complex-valued) transmission $S_{21}$
with a vector network analyzer can give access to the complex junction
admittance. However, in order to access this ideal on-chip $S$, one must
carefully calibrate the whole microwave setup using several reference devices
(e.g., thru, reflect and line) in place of the sample
{\cite{pozar_microwave_2011}} in order to deembed the effect of the rest of
the setup. Such a procedure is needed, in particular, to define a reference
for $\arg S_{21}$ and to cancel any stray transmission between input and
output ($Y_{\tmop{stray}}$ in Fig. \ref{quadrupole_model}). As our
demonstration involves evidencing only a SQUID modulation that saturates at
low temperature, it requires only qualitative measurements, and thus, for
simplicity, such a calibration is not performed. The measured (uncalibrated)
$| S_{21} |$ variations can nevertheless be qualitatively compared to the
ideal prediction
\[ | S_{21} |^2 = \frac{(2 C_c \omega Z_0)^2}{| 1 + R Y (\omega) |^2} . \]

Given our choice of parameters $R C \omega \ll 1$, the capacitive contribution
in $Y$ can be neglected for the evaluation of $S_{21}$, and we can consider
only the contribution of the Josephson element $Y (\omega) \simeq 1 / i
L_{\tmop{eff}} \omega$ (assuming a superconducting character). Under this
form, it is clear that larger values of $| S_{21} |$ correspond to small
junction admittance (large effective inductance, small supercurrent) and vice
versa.

For sample 1, in the low-temperature limit, the modulation of $| S_{21} |$
with flux is small, showing that $| Y | R \ll 1$. Assuming the junction
behaves as a usual symmetric SQUID, its inductance depends on the flux $\Phi$
as $L_{\tmop{eff}} (\Phi)^{- 1} = 2 e I_0^{\tmop{eff}} | \cos (\pi \Phi /
\Phi_0) | / \hbar$, with $I_0^{\tmop{eff}}$ the effective critical current. By
adjusting the amplitude of the $S_{21}$ modulation for sample 1 in the
low-temperature limit, this gives $I_0^{\tmop{eff}} \sim 70 \tmop{pA}$, much
smaller than the Ambegaokar-Baratoff $I_0 = E_J 2 e / \hbar = 5.0 \tmop{nA}$
value obtained from the junction's tunnel resistance. This decrease is
qualitatively expected, because zero-point phase fluctuations
{\cite{joyez_self-consistent_2013,rolland_antibunched_2019,leger_observation_2019}}
are known to reduce the effective critical current, or, equivalently, to
``renormalize'' the apparent Josephson coupling. One can also check that the
change of $| S_{21} |$ of approximately $- 1.4 \tmop{dB}$ between the maximum
frustration of the SQUID at low temperature (where $Y \simeq 0$) and the
critical temperature $T_c \sim 1.2 \text{K}$ of Al (where all lines merge at
$Y \simeq 1 / R_T$) is consistent with $R_T \simeq 62 \text{k} \Omega$, the
junction normal-state tunnel resistance. This line of reasoning is also true
for sample 2: The change of $| S_{21} |$ of approximately $- 3.6 \tmop{dB}$,
is consistent with $R_T \simeq 19 \text{k} \Omega$.

For sample 2, however, it is not possible to correctly reproduce the shape of
the variations of $S_{21}$ in the bottom left panel of Fig. \ref{exp_data} by
assuming the SQUID behaves as a standard one with $L_{\tmop{eff}} (\Phi)^{- 1}
= 2 e I_0^{\tmop{eff}} | \cos (\pi \Phi / \Phi_0) | / \hbar$ and adjusting the
effective $I_0^{\tmop{eff}}$ as done for sample 1. Given the shape of the
modulation, it seems very likely that a weak stray transmission in our setup
(as $Y_{\tmop{stray}}$ in Fig. \ref{quadrupole_model}) causes $| S_{21} |$ to
saturate at a minimum value of approximately $- 65.7 \tmop{dB}$. Note that
even if this were not the case, we expect the modulation curve could still not
be accurately fitted using $L_{\tmop{eff}} (\Phi)^{- 1} = 2 e I_0^{\tmop{eff}}
| \cos (\pi \Phi / \Phi_0) | / \hbar$, because in this sample, the $| S_{21}
|$ measurements show that the junction admittance is modulated from $| Y |
\lesssim \tmop{or} \ll 1 / R$ at the maximum frustration to $| Y | \gg 1 / R$
at minimum frustration, such that the total effective impedance at the
junction and the corresponding zero-point phase fluctuations (which determine
$I_0^{\tmop{eff}}$) vary much with $\Phi$. This variation of admittance should
lead to a strongly flux-dependent $I_0^{\tmop{eff}} (\Phi)$, and hence, an
overall non-$\tmop{abs} (\cos)$ modulation of the inverse inductance
{\cite{joyez_self-consistent_2013}}.

Finally, the striking nonmonotonic dependence of the transmission on the
temperature is explained qualitatively in the main text before the Conclusion.
\ \

Our experiment demonstrates that when quantitative predictions become
available for the junction inductance in high-impedance Ohmic environments,
calibrated $S_{21}$ measurement in such a setup should allow a quantitative
comparison.

\section{Linear response and mobility in the Caldeira-Leggett model
}\label{yvsmu}

Considering the Hamiltonian $H'$ of the main text
\[ H' = E_C (N - N_R)^2 - E_J \cos \varphi + \sum_n 4 e^2 \frac{N_n^2}{2 C_n}
   + \frac{\hbar^2}{4 e^2} \frac{\varphi_n^2}{2 L_n} . \]
the operator for the current flowing through the junction is
\[ I = \frac{2 e}{\hbar} \frac{\partial H'}{\partial \varphi} = \frac{2
   e}{\hbar} E_J \sin \varphi = I_0 \sin \varphi . \]
Now we consider a thought experiment where the junction phase $\varphi$ is
given a time dependence $\varphi \rightarrow \varphi + \delta \varphi (t)$, so
that the Hamiltonian acquires a time dependence too $H' \rightarrow H' (t)$.
We can obtain the corresponding change in the current by using the general
response formula of Ref. {\cite{safi_time-dependent_2011}} [Eq. (1) with $X
(t') \equiv \varphi (t')$ and $\hat{O} (t) \equiv I (t)$].
\begin{eqnarray}
  \frac{\delta I (t)}{\delta \varphi (t')} & = & \frac{- i}{\hbar} \theta (t -
  t') \left\langle \left[ I (t), \frac{\hbar}{2 e} I (t') \right]
  \right\rangle + \delta (t - t') \frac{2 e}{\hbar} E_J \langle \cos \varphi
  \rangle (t) \nonumber
\end{eqnarray}
where $\langle \ldots \rangle = \tmop{tr} [\rho (t) \ldots]$ with the
time-dependent density matrix $\rho (t)$. This result expresses the exact
Hamiltonian evolution, making essentially no assumption on the system or on
the drive $\delta \varphi (t)$. However, it involves the time-dependent
density matrix $\rho (t)$. The linear response of the system is obtained from
this general formula by considering vanishingly small $\delta \varphi (t)$, in
which case we can use the equilibrium density matrix in the above expression
(with the last term becoming time independent and the first one depending only
on $t - t'$).

Using the fact that the voltage fluctuations across the junction are $\delta
V = \frac{\hbar}{2 e} \frac{d}{d t} [\delta \varphi (t)]$, and going to the
frequency domain, the above result yields the junction's linear admittance
\begin{equation}
  Y (\omega) = \frac{\delta I (\omega)}{\delta V (\omega)} = \frac{1}{\hbar}
  \int_0^{+ \infty} 2 \langle [I (t), I (0)] \rangle \frac{e^{i \omega t} -
  1}{\omega} \frac{d t}{2 \pi} + \left( \frac{2 e}{\hbar} \right)^2 \frac{E_J
  \langle \cos \varphi \rangle}{i_{} \omega} \label{Yfull} .
\end{equation}
In the first term, one recognizes the standard linear susceptibility of the
usual Kubo formula. The second term is due to the (change in the) current
carried by the ground state, yielding a purely inductive response of the
junction. Even if we do not know what the equilibrium density matrix is in our
system (because the junction is entangled with the bath), this term is nonzero
as long as $E_J \langle \cos \varphi \rangle \neq 0$.

Using the fact that $2 e \dot{N} = I$, the first term in the above expression
can also be formulated in terms of the junction's charge correlator
\[ Y (\omega) = \frac{4 e^2}{\hbar} \int_0^{+ \infty} 2 \langle [N (t), N (0)]
   \rangle \omega e^{i \omega t} \frac{d t}{2 \pi} + \left( \frac{2 e}{\hbar}
   \right)^2 \frac{E_J \langle \cos \varphi \rangle}{i_{} \omega} . \]

The impedance defined by this thought experiment is an equilibrium property of
the junction coupled to its environment. In practice, when one wants to
measure this linear response, indirect driving of the junction phase can be
realized in several ways, say, by threading an ac magnetic field in the
circuit loop, or by using a capacitive bias as in our experimental setup. As
long as the probing circuitry does not alter the impedance seen by the
junction, the measured linear response is (and it {\tmem{must be}})
independent of the biasing scheme chosen. We further stress that the linear
response theory naturally embraces finite frequencies so that $Y (\omega)$ is
a genuine equilibrium property of the system, even at finite frequency. In
this regard, our probing of the system at approximately $1 \tmop{GHz}$ poses
no problem of principle.

\subsection{dc mobility vs full linear response}

In the entire literature on the Schmid-Bulgadaev transition, the transport
quantity that was focused on is the so-called dc charge mobility,
\[ \mu = \frac{\delta I_{\tmop{dc}}}{\delta V_{\tmop{dc}}} = \tmop{Re} Y
   (\omega = 0), \]
which entirely comes from the first term of the admittance (\ref{Yfull}) and
that is obtained considering only the equilibrium charge (or current)
correlator. Note that, by definition, $\mu$ describes {\tmem{dissipative}}
transport.

However, if the inductive term in $Y (\omega)$ is nonzero (i.e., if the system
can sustain a supercurrent), the zero-frequency limit of Eq. (\ref{Yfull})
considered in $\mu$ is disregarding a diverging term, and, given that there
are never strictly zero-frequency measurements, one may wonder about the
relevance of this quantity for describing transport. Indeed, we now show that
no experimental measurement protocol gives access to $\mu$ in a
superconducting system. Let us, for instance, consider the initial unbiased
equilibrium state with $\langle \varphi \rangle = 0$ in which there is no
current flowing [$I (t \leqslant 0) = 0$], and assume that at $t = 0$, a
voltage step $\delta V (t)$ is applied, ending on a plateau $\delta V \neq 0$
after a time $\tau$. If mobility were appropriately describing the linear dc
response, one would expect that after a transient, $\delta I (t \gg \tau)
\rightarrow \mu \delta V$. However, this is clearly not the case for a linear
superconducting inductor, because the inductive response to the voltage pulse
is a linearly increasing current, not a transient. It is also incorrect for a
Josephson junction and its nonlinear inductance because the interplay of the
Josephson nonlinearity and the non-Markovianity due to the retarded response
of the $R C$ circuit results in a complex dynamics of the system involving
potentially many harmonics of the Josephson frequency $\omega_J = 2 e \delta V
/ \hbar$. There is presently no general theory that is able to predict the
resulting dc current for all parameters and, in particular, when phase
fluctuations are large. Here we assume a voltage bias scheme, but one can
similarly show that the inductive response cannot be ignored in other biasing
schemes and that this cannot be fixed by changing the frequency, the
amplitude, or the temperature at which the measurement would be performed. In
a nutshell, the linear mobility simply does not properly describe transport in
a system that can sustain a supercurrent (i.e., where $E_J \langle \cos
\varphi \rangle \neq 0$ in the case we consider) because it ignores the
dominant effect of the supercurrent.

Consequently, finding a vanishing dc mobility (as in
Refs.{\cite{schmid_diffusion_1983,bulgadaev_phase_1984,guinea_diffusion_1985,aslangul_quantum_1985,zaikin_dynamics_1987,schon_quantum_1990,kimura_temperature_2004,werner_efficient_2005,lukyanov_resistively_2007}})
is not by itself a correct way of proving the system is insulating. For being
an acceptable proof, it requires, in addition, that the coherences are
suppressed in the ground state. Note that interestingly, Schmid
{\cite{schmid_diffusion_1983}} also considered the renormalization of the
coherence factor $E_J \langle \cos \varphi \rangle$ (see the following
section), but he regarded this as an {\tmem{independent}} proof of his
mobility result, and not a condition for it.

\subsection{Insulating state in the Caldeira-Leggett model? }

From the above material, it emerges that for an insulating state to exist in
this Caldeira-Leggett model, it is {\tmem{necessary}} (and sufficient) that
the environment fully suppresses the coherences between charge states in the
ground state.

As explained in the main text, RG flow analysis on the Josephson coupling
initially indicated that coherences also vanished in the insulating phase,
seemingly validating the mobility calculation (on the insulating side).
However, results on the spin-boson problem, as well as perturbation theory in
$E_J$ in the CL model contradict the RG analysis and indicate that finite
coherences always survive in the ground state of the CL model (as long as $E_J
/ E_C > 0$). Hence, from the theory point of view, it is clear by now that
within the CL model, a dissipative environment can reduce the coherence only
to a certain point. So, a remaining finite supercurrent is to be expected, and
that is indeed what we and Grimm {\tmem{et al.}} {\cite{grimm_bright_2019}}
observe. Beyond our experiment, this behavior is expected for the entire
parameter space : The junction is superconducting everywhere. This resolves
the conflicts evoked in the Introduction.

Note that our conclusion that mobility calculations do not correctly describe
transport in the CL model (and therefore cannot be used to predict a
superconducting-insulating transition) is independent of whether one considers
a compact or extended phase description; it applies also to old works which
explicitly considered an extended phase for evaluating the mobility. The
compact phase symmetry put forward by our analysis is still very important
because it enables us to reach a simple consistent picture in all known
limits, and, through the self-duality of the model, it clarifies how
decompactification occurs. \

\section{The predicted phase transition in the $P (E)$
theory}\label{tightbinding}

In Ref. {\cite{aslangul_quantum_1985}}, Aslangul {\tmem{et al.}} use a
tight-binding model to describe junctions coupled to a linear environment in
the so-called scaling limit, and they confirm Schmid's prediction of a phase
transition. Here we go over their derivation using the notations more commonly
used at present for Josephson circuits.

First we express the Hamiltonian $\tilde{H}$ considered in Ref.
{\cite{aslangul_quantum_1985}} as
\[ \tilde{H} = \frac{E_J}{2} \left( \sum_{N \in \mathbb{Z}} | N \rangle
   \langle N + 1 | e^{i \tilde{\varphi} + 2 i eVt / \hbar} + \text{H.c.}
   \right) + H_{\tmop{bath}}, \]
where $N$ is the number of charges passed through the junction and
$\tilde{\varphi}$ is the fluctuating phase across the (disconnected)
environment (in the notations of Ref. {\cite{aslangul_quantum_1985}} $B_+ =
e^{i \tilde{\varphi}}$, $\hbar \Delta = E_J$). For more generality, we
consider the case where a voltage source is present (the results of Ref.
{\cite{aslangul_quantum_1985}} are recovered taking $V = 0$). This Hamiltonian
is also considered in Ref. {\cite{ingold_charge_2005}}. Note that when using
this tight-binding description of discrete charge states, it implies the
junction phase is considered compact.

The current operator through the junction is
\[ \hat{I} = i \frac{2 e}{\hbar} \frac{E_J}{2} \left( \sum_{N \in \mathbb{Z}}
   | N \rangle \langle N + 1 | e^{i \tilde{\varphi} + 2 i eVt / \hbar} -
   \text{H.c.} \right) . \]
We now evaluate the current correlator
\[ S_{I I} (t) = \langle \hat{I} (t) \hat{I} (0) \rangle, \]
assuming that the backaction of the junction on the environment is weak enough
to not modify the equilibrium properties of the bath. At first view, this
assumption can be justified if the junction impedance at its plasma
oscillation is much larger than the environment resistance (i.e. $\sqrt{E_C /
E_J} \gg R / R_Q$), in which case the environment imposes its phase
fluctuations onto the junction. This condition is indeed fulfilled in the
scaling limit considered in Ref. {\cite{aslangul_quantum_1985}}. Within these
hypotheses, the correlator evaluates to
\begin{eqnarray}
  S_{I I} (t) & = & \left( \frac{2 e}{\hbar} \frac{E_J}{2} \right)^2 \left[
  \left( \sum_{N \in \mathbb{Z}} \langle N | \rho_{} | N \rangle \right)
  \left( \langle \text{$e^{i \tilde{\varphi} (t)} e^{- i \tilde{\varphi}
  (0)}$} \rangle e^{2 i eVt / \hbar} + \langle \text{$e^{- i \tilde{\varphi}
  (t)} e^{i \tilde{\varphi} (0)}$} \rangle e^{- 2 i eVt / \hbar} \right)
  \right.  \nonumber\\
  &  & \left. - \left( \sum_{N \in \mathbb{Z}} \langle N | \rho | N + 2
  \rangle \langle \text{$e^{i \tilde{\varphi} (t)} e^{i \tilde{\varphi} (0)}$}
  \rangle e^{2 i eVt / \hbar} + \langle N | \rho | N - 2 \rangle \langle
  \text{$e^{- i \tilde{\varphi} (t)} e^{- i \tilde{\varphi} (0)}$} \rangle
  e^{- 2 i eVt / \hbar} \right) \right]  \label{SII}
\end{eqnarray}
where $\rho$ is the reduced density matrix of the junction. Considering that
the linear environment remains in equilibrium, its fluctuations are Gaussian,
and one has
\begin{eqnarray*}
  \langle \text{$e^{\pm i \tilde{\varphi} (t)} e^{\mp i \tilde{\varphi} (0)}$}
  \rangle & = & e^{J (t)}
\end{eqnarray*}
and
\begin{eqnarray*}
  \langle \text{$e^{\pm i \tilde{\varphi} (t)} e^{\pm i \tilde{\varphi} (0)}$}
  \rangle & = & e^{- J (t) + 2 J (\infty)}
\end{eqnarray*}
with
\begin{eqnarray*}
  J (t) & = &  \langle (\tilde{\varphi} (t) - \tilde{\varphi} (0))
  \tilde{\varphi} (0) \rangle\\
  & = & \int_{- \infty}^{+ \infty} \frac{d \omega}{\omega}  \frac{\text{Re} Z
  (\omega)}{2 R_Q}  \frac{e^{- i \omega t} - 1}{1 - e^{- \beta \hbar \omega}}
\end{eqnarray*}
with $Z$ being the total environment admittance as seen from the Josephson
element, including the junction capacitance [i.e., $Z (\omega) = (R^{- 1} + i
C \omega)^{- 1}$]. For an Ohmic environment $\tmop{Re} J (\infty) = - \infty$,
so that the terms in the second line of Eq. (\ref{SII}) vanish, and using
$\tmop{tr} \rho = 1$, the correlator finally reduces to
\begin{eqnarray*}
  S_{I I} (t) & = & \left( \frac{2 e}{\hbar} \frac{E_J}{2} \right)^2 e^{J (t)}
  \cos \frac{2 e V t}{\hbar},
\end{eqnarray*}
or, in the frequency domain,
\[ \begin{array}{lll}
     S_{I I} (\omega) & = & \left( \frac{2 e}{\hbar} \frac{E_J}{2} \right)^2
     (P (\hbar \omega + 2 e V) + P (\hbar \omega - 2 e V)) .
   \end{array} \]
where
\[ \begin{array}{lcl}
     P (E) & = & \frac{1}{2 \pi \hbar}  \int_{- \infty}^{\infty} \exp [J (t) +
     iEt / \hbar] dt
   \end{array} \]
is the usual $P (E)$ function considered in dynamical Coulomb blockade. For
the $RC$ environment considered here, at zero temperature one has
{\cite{ingold_charge_2005}}
\begin{equation}
  P (E) \propto E^{2 R / R_Q - 1} \label{POhmicT0} .
\end{equation}

\subsection{Charge transport}

The {\tmem{standard}} Green-Kubo relations link the admittance $Y_{\tmop{GK}}$
to $S_{I I}$
\begin{equation}
  \tmop{Re} Y_{\tmop{GK}} (\omega, V) = \frac{1}{2 \omega} [S_{I I} (\omega) -
  S_{I I} (- \omega)] \label{Ykubo} .
\end{equation}
Note that, even after applying Kramers-Kronig relations to get the imaginary
part, this admittance $Y_{\tmop{GK}}$ corresponds only to the first term in
Eq. (\ref{Yfull}) and therefore lacks the inductive response of the junction,
which we know is important (see Appendix \ref{yvsmu}). If we nevertheless
proceed, from Eq. (\ref{Ykubo}) one predicts a differential conductance
\begin{eqnarray*}
  \frac{d I}{dV} (V) & = & \tmop{ReY}_{\tmop{GK}} (\omega \rightarrow 0, V)\\
  & = & \frac{2 e^2}{\hbar} E_J^2 [P' (2 e V) + P' (- 2 e V)]
\end{eqnarray*}
and the $I \text{-} V$ characteristics are obtained by straightforward
integration
\begin{equation}
  I (V) \begin{array}{ll}
    = & \frac{e}{\hbar} E_J^2 [P (2 e V) + P (- 2 e V)] .
  \end{array} \label{naiveIV}
\end{equation}
The above results are already found in Ref. {\cite{ingold_charge_2005}}. They
describe {\tmem{inelastic}} tunneling processes of Cooper pairs with real
transitions in the environment modes. The $I \text{-} V$ characteristic
(\ref{naiveIV}) is known to quantitatively describe experiments
{\cite{hofheinz_bright_2011,rolland_antibunched_2019}} at finite voltages when
the Josephson coupling is small enough that the environment modes remain in
equilibrium. \

From the above results one predicts the junction zero-bias conductance
\begin{eqnarray}
  G & = & Y_{\tmop{GK}} (\omega \rightarrow 0, V = 0) \nonumber\\
  & = & \frac{2 e}{\hbar} E_J^2 P' (0)  \label{Gkubo}
\end{eqnarray}
which corresponds to the dc charge mobility calculated by Aslangul {\tmem{et
al.}} {\cite{aslangul_quantum_1985}}. Thus, for the $RC$ environment
considered here, at $T = 0$, using Eqs. (\ref{POhmicT0}) and (\ref{Gkubo}),
one recovers the ``superconducting-to-insulating'' phase transition at $R =
R_Q$, as found by Aslangul {\tmem{et al.}} {\cite{aslangul_quantum_1985}}
(Schmid {\cite{schmid_diffusion_1983}} and Bulgadaev
{\cite{bulgadaev_phase_1984}} obtained the same results for the mobility by
mapping the problem onto a log-gas). However, as noted above, the
charge-transfer processes described here are inelastic and it is therefore not
correct to describe this type of process as {\tmem{superconducting}} transport
for $R < R_Q$.

\subsection{Conclusions on the phase transition}

\begin{enumerateroman}
  \item As shown in Appendix \ref{yvsmu}, $Y_{\tmop{GK}}$ is not the full
  linear admittance; it entirely misses the inductive response of the junction
  and cannot properly describe charge transport (and notably the supercurrent
  branch in the {\tmem{I-V}} characteristics) because of that. This
  perturbative tight-binding approach does predict a transition in the charge
  correlator (a partial charge localization), but it is incorrect to infer
  from this result that a superconducting-insulating transition exists.
  
  \item The predicted transition in the junction charge correlator arises
  entirely from the $P (E)$ function, i.e. from the equilibrium fluctuations
  across the $RC$ environmental impedance not connected to anything. It has
  nothing to do with the junction. This is already noted in Ref.
  {\cite{guinea_diffusion_1985}}.
  
  \item As we mention in the main text, for $R < R_Q$, the charge is
  predicted to be fully delocalized, and correspondingly the junction phase is
  fluctuationless (and its dynamics is that of a classical quantity). 
\end{enumerateroman}

The last two points seem odd and most likely too sketchy. Just as it is now
understood that the coherence factors do not actually vanish in this system,
it is quite clear that taking into account the backaction of the junction on
the environment (causing their entanglement) would suppress the above sharp
transition in the charge correlator and turn it into a smooth crossover with
finite but small phase (resp. charge) fluctuations in the (resp. dual of the)
delocalized charge state. Actually, we know this is the expected behavior when
$E_J \gg E_C$ and $R \gg R_Q$ [upper left corner of the diagram in Fig.
\ref{new_diagrams}(b)]: At low temperatures, such a junction behaves
essentially as a linear inductor and it is well known that parallel $R L C$
circuits have finite charge and phase fluctuations for all parameters. Then,
using the duality argument, the presence of finite charge and phase
fluctuations should also be true for $E_J \ll E_C$ and $R < R_Q$. Finally, by
continuity, this should be also true in the entire diagram.

The continuous crossover that emerges from our analysis contrasts with the
results of the Monte Carlo simulations performed assuming an extended phase in
Ref. {\cite{werner_efficient_2005}}, where an abrupt transition in the phase
correlator at $T = 0$ and $R = R_Q$ is found. This discrepancy illustrates
that considering an extended phase can lead to results inconsistent with our
analysis (see Appendix \ref{compact-extended}).

In conclusion, we expect no DQPT transition in the CL model: neither a
superconducting-insulating transition nor a transition in the charge or phase
correlator.

\section{Compact vs Extended Junction Phase}\label{compact-extended}

The analysis of the CL model conducted in the main text is based on symmetry
considerations and leads to a ``phase diagram'' [Fig. \ref{new_diagrams}(b)],
which is theoretically consistent (including at its boundaries) and consistent
with experiments. In this phase diagram, the junction phase is compact below
the antidiagonal and it progressively decompactifies above the antidiagonal,
in a smooth crossover.

We stress that when this decompactification does not occur, compact phase
solutions are dictated by the symmetry of (the effective Josephson Hamiltonian
in) the CL model; this symmetry is {\tmem{not}} for the physicist to
{\tmem{choose}}. As a corrollary, {\tmem{choosing}} to use an extended phase
in many earlier works on the CL model cannot be rigorously justified
theoretically because there are no known mechanisms within that model that
would break the system's fundamental symmetry in this way. The only
established symmetry breaking for the phase is the partial decompactification
we describe in the main text (but it cannot be found starting from an extended
phase).

Since it cannot be justified within the model, making use of an extended
phase implicitly and forcefully adds poorly controlled hypotheses or
ingredients to the model, with essentially unknown consequences (In practice,
it adds an additional variable indexing the wells of the cosine and enabling
to distinguish all of them in all circumstances, which is not possible in the
original model.) For sure, this can be done theoretically --- it works. But
does such a treatment still yield fully relevant predictions for the
real-world system that the model was originally meant to describe? Clearly,
not. It is certain that predictions will differ in circumstances where
interferences between the wells matter, and this difference is unavoidable in
a system with superconducting coherences such as the one we consider. In other
words, in the parameter space where we now know the phase discrete
translational symmetry is not spontaneously broken, there exists
mathematically correct extended phase solutions for $H$ that cannot be
unitarily transformed to a suitable (i.e., compact) solution for $H'$. Such
solutions do not respect the intrinsic system symmetry, but it is nearly
impossible to figure this inconsistency by considering only the extended phase
hypothesis.

\

This subtle point on the junction phase symmetry and its spontaneous breakage
has never been properly understood so far. We think that bringing this point
up and clarifying it is a significant achievement of the present work.

\subsection{Retrospective on the compact vs extended phase debate }

Prior to this work, it was intuited that phase decompactification must take
place somehow (at least for some parameters) but it was not understood how it
was occurring and this resulted in a lot of ambiguities and confusion. Here,
we try to put into perspective why the situation was so confusing.

An extended phase description contains the compact phase solutions as
solutions of higher symmetry (periodic solutions in phase representation), so
that, in principle, it should be the only description ever needed. Indeed,
when starting from a Hamiltonian such as $H$ in the main text, for which an
extended phase is the ``natural'' point of view, one can obtain the compact
phase solutions by considering highly nontrivial initial and boundary
conditions {\cite{loss_effect_1991-1,mullen_resonant_1993}}. However, in the
existing literature based on using $H$, this was not done and, as a
consequence, compact phase solutions (which are of utmost importance as our
work shows) were not found or not recognized as such.

Until now, this seemed not too problematic and it was even rationalized that
compact phase states were irrelevant in systems that are most conveniently
described using $H$ (essentially systems where a dc current can flow). The
rationale was that in these systems, a ``full decompactification'' process
(i.e., yielding only nonperiodic extended phase states solutions of $H$) would
always occur for all parameters and all temperatures. At first, this was just
argued for qualitatively {\cite{averin_bloch_1985}}. Soon after, Zwerger
{\tmem{et al.}} {\cite{zwerger_effects_1986}} showed that such a full
decompactification process should indeed always occur for Ohmic environments,
but their derivation can no longer be considered conclusive as it did not take
into account the entanglement of the junction with the environment, which we
now know is key. Later, Apenko proposed another justification
{\cite{apenko_environment-induced_1989}}, but in his derivation, the
identification of different phases in the circuit was not rigorous (similar to
what we discuss about the Hamiltonian of the fluxonium circuit in Appendix
H\ref{fluxonium}).

Hence, schematically, for a very long time, it was broadly considered that
the symmetry of the phase and the Hamiltonian used were somehow tied : $\left(
H \Longleftrightarrow \tmop{extended} \tmop{phase} \text{assumed to be a
decompactified phase} \right)$ XOR $(H' \Longleftrightarrow \tmop{compact}
\tmop{phase})$.

To support this dichotomic view, several arguments or criteria were used to
favor using a compact or an extended phase description, depending on the
problem considered. For instance it was frequently argued that a compact
junction phase is suitable only in circuits having an ``island'' connected to
the junction as it would be a manifestation of the charge quantization in the
island or of the tunneling of individual Cooper pairs through the junction. In
other words, a compact phase should not be appropriate in a circuit where the
charge can flow continuously (and thus considering $H'$ to describe the Ohmic
shunted junction was not considered appropriate). Although the general
discussion of the main text already shows such arguments are not relevant, in
the following subsections of this appendix we nevertheless specifically
discuss why these arguments do not hold.

\subsection{Phase compactness is not due to the tunneling of individual
Cooper pairs through the junction}

If instead of a Josephson junction one considers a superconducting ballistic
(or nearly ballistic) weak link, then the current-phase relation is still
periodic with the phase, so that one can again use a discrete charge basis to
describe the state of the weak link. In that case, this apparent ``charge
discretization'' obviously cannot be directly linked to an underlying charge
quantization due to the tunneling of charge carriers.

\subsection{Is charge quantization due to the presence of islands?}

As we discuss in the main text, using the discrete charge basis of the CPB
(equivalent to considering a compact phase) arises from the symmetries of the
system. It does not require the presence of ``an island'' in which the charge
is ``naturally quantized.'' The simplest argument against this is that in a
CPB the mere presence of the Josephson junction destroys this charge
quantization (the ground state of the CPB consists of a coherent superposition
of charge states, with finite zero-point fluctuations). This ``charge
quantization'' is not observable, it is only a mathematical illusion,
actually.

Our statement is further supported by the fact that the form of the
Caldeira-Leggett Hamiltonian is independent of whether the circuit has an
island or not. This can be shown using the explicit decomposition of the total
circuit impedance into oscillators according to the rules in Ref.
{\cite{vool_introduction_2017}}.

Finally one can show that the Hamiltonian of a circuit with an island has a
smooth limit to the islandless case by taking the limit where the capacitance
defining the island becomes infinite. Correspondingly, all the
finite-frequency linear response functions of the system have smooth limits
too. However, as the system is nonlinear, the linearity range may vanish at
low frequency (see, e.g., Appendix \ref{former_exps}), depending on the type
of response probed. This agrees with the obvious expectation that at strictly
zero frequency no dc current can flow when there is an island, while it can if
there is no island. As we explain in the main text, the absence of dc current
in a circuit with an island results from having a single ground state, while
there is a continuum of them in the islandless case permitting a dc current
flow.

As a conclusion, whether one considers a CPB with an island or a galvanically
shunted junction does not radically change the way the system is modeled.

\subsection{The junction's phase in the fluxonium}\label{fluxonium}

It is frequently argued that one {\tmem{must}} use an extended phase
description for describing the fluxonium circuit
{\cite{manucharyan_fluxonium_2009-2}} where a Josephson junction is connected
in parallel with a inductor (instead of a resistor in this paper).

Indeed, for the fluxonium, the Hamiltonian proposed in Refs.
{\cite{manucharyan_fluxonium_2009-2,koch_charging_2009}} is \

\begin{equation}
  H_{f 1} = \frac{q_{}^2}{2 C} - E_J \cos \varphi_{} + \frac{\left(
  \Phi_{\tmop{ext}} - \frac{\hbar}{2 e} \varphi_{} \right)^2}{2 L} \label{Hf1}
\end{equation}
where $\frac{\hbar}{2 e} \varphi_{}$ and $q$ denote the branch flux and charge
of the junction and $\Phi_{\tmop{ext}}$ is the magnetic flux enclosed by the
loop formed between the junction and the inductor considered as an external
control parameter, i.e., a fixed real number. In this model, obviously not
invariant upon $\varphi \rightarrow \varphi + 2 \pi$, the junction's phase
clearly appears as extended. However, the eigenstates of the system have
current fluctuations that, in addition to vacuum flux fluctuations, cause
fluctuations of the flux in the loop, which contradicts the assumption that \
$\Phi_{\tmop{ext}}$ is a fixed parameter. Thus, the model is not fully
consistent.

Another fluxonium Hamiltonian is derived in Ref. {\cite{ulrich_dual_2016-1}}.
It reads
\begin{equation}
  H_{f 2} = \frac{(Q + q)_{}^2}{2 C} - E_J \cos \varphi_{} +
  \frac{\Phi_{}^2}{2 L} \label{Hf2} .
\end{equation}
In this writing, $\Phi_{}$ and $Q$ denote the branch flux and charge of the
inductor, while $\frac{\hbar}{2 e} \varphi_{}$ and $q$ still denote the branch
flux and charge of the junction. This Hamiltonian thus has 2 quantum degrees
of freedom (each with fluctuations), and the flux in the loop is given by
Kirchhoff's law
\[ \frac{\hbar}{2 e} \varphi - \Phi = \Phi_{\tmop{loop}} \]
so that $\Phi_{\tmop{loop}}$ fluctuates too (as expected) and has an
expectation value related to the externally applied flux $\Phi_{\tmop{ext}}$.
It is only by suppressing one of the quantum degrees of freedom, turning it
into a classical one, that Eq. (\ref{Hf2}) becomes Eq. (\ref{Hf1}) (and,
strictly, $\varphi$ can no longer be considered as a degree of freedom
describing the sole junction). The junction's phase appearing as extended in
Eq. (\ref{Hf1}) thus results from an approximation (perhaps a very good one);
it is not an obligation.

The inconsistency pointed out above is a general problem of the circuit
quantization scheme proposed in Ref. {\cite{vool_introduction_2017}}, where
loop fluxes are always assumed constant. It can be easily fixed though. Other
quantization schemes have also been proposed
{\cite{burkard_multilevel_2004,burkard_circuit_2005,ulrich_dual_2016-1}} which
do not necessarily force this approximation.

\

\paragraph{The fluxonium is not in the phase diagram.}

In the fluxonium circuit, the impedance seen by the junction has $\tmop{Re} Z
(\omega = 0) = 0,$ which would naively locate it on the right axis of the SB
phase diagram. However, in that limit, the system considered in the main text
is ill-defined as neither the loop inductance $L$ (which defines a new energy
scale $E_L = \hbar^2 / 8 e^2 L$ in the problem) nor the external flux
$\Phi_{\tmop{ext}}$ threading the loop are specified. Thus the phase diagram
would need to be refined with extra parameters close to the right axis.

Nevertheless, depending on its parameters, we expect the fluxonium's junction
phase will evolve between fully decompactified (in a single well of the
cosine) when $E_L \gg E_J$ and $\Phi_{\tmop{ext}} \tmop{mod} \Phi_0 \neq
\frac{1}{2}$, partially decompactified (in several wells) when $E_L \sim E_J$
and essentially compact (populating many wells nearly equally)
{\cite{koch_charging_2009}} when $E_L \rightarrow 0$.

\subsection{Phase in current-biased junctions}

When considering the case of a current-biased junction, where the current
source ``tilts the washboard potential'', the different wells of the cosine
appear as nonequivalent. Here again, the obligation to use an extended phase,
is only apparent.

First, the current source can be modeled by considering a very large inductor
loaded with an initial flux. So we are back to considering the fluxonium case
for which we argue above that there is no obligation to use an extended phase.

One can arrive at a similar conclusion by performing a time-dependent unitary
transformation {\cite{loss_effect_1991-1}} that removes the tilt of the
washboard, restoring the periodicity of the cosine potential. In this case,
however, the states of the system will be time dependent.

In such a current-biased junction, the final degree of phase
decompactification will depend on the dissipation in the system and on the
ratio $E_J / E_C$ (as in the unbiased case), but certainly also on the current
bias $I_b$ which sets an extra energy scale $I_b \Phi_0$ in the system, with
an associated dynamics.

\section{Validity of the effective Josephson Hamiltonian and consistency of
the Caldeira-Leggett model}\label{effectiveH}

In the CL model, the junction is modeled using the effective Josephson
Hamiltonian ( i.e. the celebrated washboard potential for the junction phase)
which describes only Cooper pair tunneling and one couples this effective
Hamiltonian to the linear environment.

This effective Josephson Hamiltonian emerges from the tunneling of
quasiparticles at second order in perturbation theory in absence of an
environment {\cite{ambegaokar_tunneling_1963,ambegaokar_quantum_1982}} and it
is commonly admitted it describes well a junction at energies much lower than
the superconducting gap $\Delta$ and in the absence of quasiparticles (which
is expected at $k_B T \ll 2 \Delta$). Even when these conditions are
fulfilled, one may wonder whether considering the effect of the environment on
this effective Hamiltonian ---as done in the CL model--- is fully consistent.

A more rigorous and consistent way of considering the effect of the
environment on the junction consists of going back to the tunneling of
quasiparticles
{\cite{martinis_energy_2009-1,joyez_self-consistent_2013,ansari_effect_2013}}.
Doing so, one however finds that at $\tmop{second}$ order in tunneling
(corresponding to the effective Josephson Hamiltonian used in $H$ or $H'$) the
junction sees the bare zero-point fluctuations of the $R C$ circuit. However,
at that lowest order in perturbation, phase fluctuations are divergent for any
Ohmic environment and this divergence predicts a complete suppression of the
supercurrent at all temperatures {\cite{joyez_self-consistent_2013}}, even for
$R < R_Q$, in the phase where a classical compact phase is predicted. This
shows that the CL description of the system (using $H$ or $H'$) is
inconsistent when considering an Ohmic environment. These inconsistencies
resolve at higher orders in the tunneling Hamiltonian (or using a
self-consistent approximation {\cite{joyez_self-consistent_2013}}), when the
inductive backaction of the junction on the environment is taken into account
: The junction and environment become entangled, voltage and phase
fluctuations are reduced, and they acquire an effective super-Ohmic spectral
density for which no DQPT is expected, the junction preserving a finite
supercurrent at $T = 0$ for all environmental impedances.

Our present work shows that even within the CL model (although it is not fully
consistent), the predicted phase transition similarly disappears when
considering the backaction of the junction on the environment.

\section{Relationship with phase transitions in other systems
}\label{otherQPTs}

The phase transition predicted by SB is closely related to a number of other
phase transitions predicted in different systems (see Ref.
{\cite{torre_viewpoint_2018}}).

In particular it is related to the impurity-induced transition in a 1D
conducting channel of interacting spinless fermions (i.e. a Tomonaga-Luttinger
liquid, TLL) predicted by Kane and Fisher {\cite{kane_transport_1992}} (KF),
according to which, at $T = 0$, for any nonzero strength of the impurity
potential, the channel conductance should vanish for repulsive interactions $g
< 1$, while it should reach the perfect TLL conductance $g e^2 / h$ for
attractive interaction $g > 1$. This behavior is akin to the SB prediction of
a superconducting-to-insulating transition. Kane and Fisher showed that these
systems are indeed described by the same effective action and, according to
the principle ``the same equations have the same solutions'' made famous by
Feynman, no one questioned they would have the same phase transition physics
until now, even when it became evident that the SB prediction conflicted with
known results on Josephson junctions.

\subsection{Confirmations of the KF phase transition}

Repulsive Luttinger liquids with rational values of $g < 1$ have been
extensively studied theoretically since they notably describe the low-energy
physics of fractional quantum Hall edge states
{\cite{wen_electrodynamical_1990}}. Thanks to the methods of integrable
systems, exact results have been obtained for the specific values of $g = 1 /
2$ {\cite{kane_transmission_1992}}, $g = 1 / 3$ {\cite{fendley_exact_1995}}
and $g = 2 / 3$ {\cite{anthore_circuit_2018}}. All these results corroborate
the perturbative RG analysis {\cite{kane_transport_1992}} predicting universal
scaling laws for the dc conductance which drive the system to an insulating
state as the temperature is lowered, for all impurity backscattering strength.

The KF phase transition physics was confirmed experimentally by taking
advantage of a second mapping put forward by Safi and Saleur
{\cite{safi_one-channel_2004-1}}, who noticed the action of an impurity in a
TLL is also equivalent to that of a single-channel quantum point contact in
series with a resistor (QPC+$R$). In this mapping the TLL interaction
parameter $g$ is controlled by the resistance $g = 1 / (1 + R / 4 R_Q)$, and
thus covers only the dynamics of repulsive TLLs. Since the physical
implementation of QPC+$R$ is much better controlled than that of fractional
quantum Hall physics, this mapping enabled precise experimental investigations
of the dc-conductance scaling laws. The experiments reported in Refs
{\cite{jezouin_tomonagaluttinger_2013-1,parmentier_strong_2011-1,anthore_circuit_2018}}
provide stringent tests of the predicted universal critical behavior at low
energies (temperature and dc voltage) even though the system is not strictly
in the scaling limit because of the finite charging energy.

\subsection{Same equations but different solutions?}

At first sight it is quite shocking that we invalidate the SB phase transition
after the KF one was accurately confirmed; it obviously violates ``the same
equations have the same solutions'' principle.

The key of this paradox is that the principle makes implicit assumptions on
the equations' {\tmem{context}}. Everyone knows a given real-coefficient
polynomial $p (x)$ may have roots or not depending whether the context is $x
\in \mathbb{C}$ or $x \in \mathbb{R}$. The SB and KF systems can be described
by the same effective action, but when one goes back to the underlying
microscopic descriptions, different phenomenologies arise, providing different
contexts for searching the solutions to the equations.

For a Josephson junction, there is a gap in the excitation spectrum of its
electrodes. Consequently, after a slow enough $2 \pi$ phase slip, the junction
is still in its ground state, and since the initial and final states of the
junction are indiscernible, the junction's phase is compact. Superconductivity
also yields a static phase coherence $\langle \cos \varphi \rangle \neq 0$,
and an inductive response. Our work shows that this ``superconducting
context'' is robust to connecting a resistor to the junction: The junction and
the bath entangle, preserving finite coherences which forbids the phase
transition. To put it more simply, in circuit engineering terms, the
superconducting (ground-state) inductive response shunts the low-frequency
phase fluctuations arising from the series resistance; this makes the global
system super-Ohmic, allowing the junction to preserve its superconducting
character.

No such mechanism can take place in KF or QPC+$R$ systems. In the case of an
open 1D electonic channel with a barrier connected at both ends to reservoirs,
a $2 \pi$ phase slip at the barrier (however slow) corresponds to a voltage
pulse which, at $T = 0$, can excite electrons and/or holes at arbitrarily low
energy and which will be dissipated in the reservoirs. Thus, a phase slip
takes this system to an orthogonal (distinguishable) state, such that the
phase needs to be regarded as an extended phase. The Fermionic baths hence
provide a subtle mechanism that is not contained in the equations of the
effective model (where the Fermions no longer appear) and that allows breaking
the discrete translational invariance of the phase in a way that totally
differs from the partial decompatification mechanism we identify in JJs.
Furthermore, in this system there is no possibility of a supercurrent in the
ground state, and thus no static coherence ($\langle \cos \varphi \rangle =
0$). Connecting a resistor to the channel brings the system in the critical
regime of the DQPT, with the expected localization effect described by Schmid
and Bulgadaev.

The above discussion suggests that the compactness of the phase ---which we
justify in the main text from the symmetry of the effective Josephson
Hamiltonian--- cannot be detached from the superconducting character of the
Josephson junction and the existence of the inductive response.

\subsection{Superconducting-to-insulating transition in 1D JJ arrays}

Another superconducting-to-insulating phase transition is predicted in 1D JJ
arrays {\cite{bradley_quantum_1984}}. This latter transition was related to
the disordered-induced transition (i.e. Anderson localization) predicted in
fermionic 1D systems {\cite{giamarchi_anderson_1988}}. Recently, Kuzmin
{\tmem{et al.}} investigated experimentally 1D JJ arrays and observed they
remained good superconducting transmission lines well beyond the threshold
line impedance predicted for their transition to the insulating state
{\cite{kuzmin_quantum_2019-1}}. Given the similarities between that system and
the one we consider, we believe it could be worth revisiting the predicted
transition in 1D JJ arrays taking into account what we understood on the
sensitivity of the SB transition to the superconducting character of the
underlying system.


\begin{thebibliography}{10}
\bibitem[32]{note3}Note that in standard CPB circuits where the junction is
  isolated from the dissipative environement by a gate capacitor not allowing
  a dc charge flow, this translational invariance of the charge is suppressed
  and the quasicharge is a relevant parameter.
\end{thebibliography}
\end{document}